\documentclass[iop]{emulateapj}
\usepackage{amsmath}
\usepackage{tabularx}
\usepackage{changepage}

\begin{document}
\title{Synthesizing Exoplanet Demographics from Radial Velocity and Microlensing Surveys, II: The Frequency of Planets Orbiting M Dwarfs}

\author{Christian Clanton, B. Scott Gaudi}
\affil{Department of Astronomy, The Ohio State University, 140 W. 18th Ave., Columbus, OH 43210, USA}
\email{clanton@astronomy.ohio-state.edu}

\begin{abstract}
In contrast to radial velocity surveys, results from microlensing surveys indicate that giant planets with masses greater than the critical mass for core accretion ($\sim 0.1~M_{\rm Jup}$) are relatively common around low-mass stars. Using the methodology developed in the first paper, we predict the sensitivity of M-dwarf radial velocity (RV) surveys to analogs of the population of planets inferred by microlensing.  We find that RV surveys should detect a handful of super-Jovian ($>M_{\rm Jup}$) planets at the longest periods being probed. These planets are indeed found by RV surveys, implying that the demographic constraints inferred from these two methods are consistent. We show that if total RV measurement uncertainties can be reduced by a factor of a few, it is possible to detect the large reservoir of giant planets ($0.1-1~M_{\rm Jup}$) comprising the bulk of the population inferred by microlensing. We predict that these planets will likely be found around stars that are less metal-rich than the stars which host super-Jovian planets. Finally, we combine the results from both methods to estimate planet frequencies spanning wide regions of parameter space.  We find that the frequency of Jupiters and super-Jupiters ($1\lesssim m_p\sin{i}/M_{\rm Jup}\lesssim 13$) with periods $1\leq P/{\rm days}\leq 10^4$ is $f_{\rm J}=0.029^{+0.013}_{-0.015}$, a median factor of 4.3 ($1.5-14$ at 95\% confidence) smaller than the inferred frequency of such planets around FGK stars of $0.11\pm 0.02$. However, we find the frequency of all giant planets with $30\lesssim m_p\sin{i}/M_{\oplus}\lesssim 10^4$ and $1\leq P/{\rm days}\leq 10^4$ to be $f_{\rm G}=0.15^{+0.06}_{-0.07}$, only a median factor of 2.2 ($0.73-5.9$ at 95\% confidence) smaller than the inferred frequency of such planets orbiting FGK stars of $0.31\pm 0.07$. For a more conservative definition of giant planets ($50\lesssim m_p\sin{i}/M_{\oplus}\lesssim 10^4$), we find $f_{\rm G'}=0.11\pm 0.05$, a median factor of 2.2 ($0.73-6.7$ at 95\% confidence) smaller than that inferred for FGK stars of $0.25\pm 0.05$. Finally, we find the frequency of all planets with $1\leq m_p\sin{i}/M_{\oplus}\leq 10^4$ and $1\leq P/{\rm days}\leq10^4$ to be $f_p=1.9\pm 0.5$.
\end{abstract}


\section{Introduction}
The ever-increasing number of exoplanet discoveries has enabled the characterization of the underlying population of planets in our galaxy. Planet frequencies have been determined by multiple detection methods: RV \citep[e.g.][]{2005ApJ...622.1102F,2008PASP..120..531C,2008A&A...487..373S,2009A&A...493..639M,2010PASP..122..905J,2010Sci...330..653H,2011arXiv1109.2497M,2013A&A...549A.109B}, transits \citep{2006ApJ...644L..37G,2011ApJ...736...19B,2011ApJ...742...38Y,2011ApJ...738..151C,2012ApJS..201...15H,2012ApJ...745...20T,2013ApJ...764..105S,2013ApJ...767...95D,2013ApJ...766...81F}, microlensing \citep{2002ApJ...566..463G,2010ApJ...720.1073G,2010ApJ...710.1641S,2011Natur.473..349S,2012Natur.481..167C}, and direct imaging \citep{2010ApJ...717..878N,2011ApJ...733..126C,2012A&A...541A.133Q}. These studies have provided interesting results, but, individually, are constrained to limited regions of parameter space (i.e. some given intervals of planet mass and period). Synthesizing detection results from multiple methods to derive planet occurrences that cover larger regions of parameter space would provide much more powerful constraints on demographics of exoplanets than is provided by individual techniques. Such synthesized data sets will better inform formation and migration models of exoplanets.

Perhaps surprisingly, M dwarf hosts are the best characterized sample in terms of exoplanet demographics. RV surveys are most sensitive to planets on orbits smaller than a few AU (ultimately depending on the duration and cadence of a given survey). At large separations, from $\sim 10$ to $100~$AU, direct imaging is currently the only technique with the capability to provide information, and then, only for young stars. The only method capable of deriving constraints on the demographics of exoplanets in the intermediate regime of separations from a few to $\sim 10~$AU is microlensing. However, for a range of lens distances, $dD_l$, the contribution to the rate of microlensing events scales as $\propto n\left(D_l\right)M_l^{1/2}$, where $n\left(D_l\right)$ is the number density of lenses and $M_l$ is the lens mass. Thus, the integrated microlensing event rate is explicitly dependent on the mass function of lenses. The slope of the mass function for $M_l\lesssim M_{\odot}$ is such that there are roughly equal numbers of lens stars per logarithmic interval in mass. Thus, lower mass objects are more numerous and more often act as lenses in a microlensing event. Indeed, \citet{2010ApJ...720.1073G} (hereafter GA10) report the typical mass in their sample of microlensing events to be $\sim 0.5~M_{\odot}$. This means that constraints on exoplanet demographics at ``intermediate'' separations (few to $\sim 10~$ AU) exist primarily for M dwarfs, as that is the population best probed by microlensing.

The low giant planet frequencies around M dwarfs inferred from RV surveys have been been heralded as a victory for the core accretion theory of planet formation, which makes the generic prediction that giant planets should be rare around such stars \citep{2004ApJ...612L..73L,2005ApJ...626.1045I,2008ApJ...673..502K}. However, microlensing has found an occurrence rate of giant planets, albeit planets that are somewhat less massive than those found by RV (but nevertheless still giant planets), that is more than an order of magnitude larger than that inferred from RV. On the other hand, microlensing is sensitive to larger separations than RV, typically detecting planets beyond the ice line. If the microlensing results are correct, they imply that giant planets do form relatively frequently around low mass stars, but do not migrate, perhaps posing a challenge to core accretion theory.

Table~\ref{tab:freq_constraints} lists the constraints on giant planet occurrence rates around M dwarfs from the microlensing survey of GA10 and the RV surveys of \citet{2010PASP..122..905J} (hereafter JJ10) and \citet{2013A&A...549A.109B} (hereafter BX13), including the planetary mass and orbital period intervals over which the frequency measurements are valid.

\begin{table*}
\centering
\caption{\label{tab:freq_constraints} Planet frequency around M dwarfs from microlensing and RV surveys. The mass and period intervals for the microlensing measurement were estimated using the typical lens mass of $M_l\sim 0.5~M_{\odot}$ and the typical mass ratio $q\sim 5\times10^{-4}$. The mass limit for the CPS sample assumes a $0.5~M_{\odot}$ host at an orbital separation of 1~AU. See \S~\ref{sec:sample_properties} for details.}
\begin{tabular}{@{\extracolsep{0pt}}lcccc@{\extracolsep{0pt}}}
\hline
\hline
\rule{0pt}{2.6ex}\rule[-1.8ex]{0pt}{0pt} & $\frac{d^2N}{d\log{\left(m_p\sin{i}\right)}d\log{\left(a\right)}}$ $\left[{\rm dex^{-2}}\right]$& Period Interval [days] & Mass Interval $\left[M_{\oplus}\right]$ &  Reference \\ \hline
\hline
 Microlensing & $0.36\pm 0.15$ & $560\lesssim P \lesssim 5600$ & $10\lesssim m_p\sin{i} \lesssim 3000$ & GA10 \\ \hline
 HARPS (RV) & $0.0080^{+0.0077}_{-0.0043}$ & $P \lesssim 2000$ & $m_p\sin{i}\gtrsim100$ & BX13 \\ \hline
 CPS (RV) & $0.0085\pm 0.0041$ & $P \lesssim 2000$ & $m_p\sin{i}\gtrsim150$ & JJ10 \\ \hline
\end{tabular}
\end{table*}

There are several potential reasons for this large difference in inferred giant planet frequency. The properties and demographics of the observed sample of host stars observed with microlensing could well be different from the targeted (local) M dwarfs monitored with RV. RV studies have shown a clear trend of planet occurrence with metallicity \citep{2005ApJ...622.1102F,2010PASP..122..905J,2013A&A...551A..36N,2014ApJ...781...28M} and the slope of the Galactic metallicity gradient \citep[see e.g.][and references therein]{2012ApJ...746..149C,2013arXiv1311.4569H} suggests that the metallicity distribution of local M dwarfs is systematically lower than that of the GA10 microlensing sample. Furthermore, some of the lenses in the GA10 microlensing sample could be K or G dwarfs, or even stellar remnants, although the fraction of events with such lenses to all events is expected to be relatively low \citep[e.g.][]{2000ApJ...535..928G}. It could also be that the population of planets orbiting local M dwarfs differs from the population orbiting M dwarfs in other parts of the galaxy, and in particular, planets orbiting stars in the Galactic bulge.

However, perhaps the simplest potential explanation for the large discrepancy in the observed giant planet frequency around M dwarfs is the different ranges of orbital period and planet mass probed by the two discovery methods. Indeed, \citet{clanton_gaudi14a} suggest that the slope of the planetary mass function is sufficiently steep that even a small difference in the minimum detectable planet mass can lead to a large change in the inferred frequency of planetary companions.

Thus, motivated by the order-of-magnitude difference in the frequency of giant planets orbiting M dwarfs inferred by microlensing and RV surveys, we have developed in a companion paper \citep{clanton_gaudi14a} the methodology necessary to statistically compare the constraints on exoplanet demographics inferred independently from these two very different discovery methods. We also justify the need for a careful statistical comparison between these two datasets by showing an order of magnitude estimate of the velocity semi-amplitude, $K$, and the period, $P$, of the ``typical'' microlensing planet, which we define as one residing in the peak region of sensitivity for the GA10 microlensing sample. This typical planet has a host star mass of $M_l\sim 0.5~M_{\odot}$, a planet-to-star mass ratio of $q\sim 5\times10^{-4}$, and a projected separation of $r_{\perp}\sim 2.5~$AU, corresponding to a planet mass of $m_p \sim 0.26~M_{\rm Jup} \sim{\rm M_{\rm Sat}}$. We find that for $\sin{i}\approx 0.866$ (the median value for randomly distributed orbits) and a circular orbit, the typical microlensing planet will have a period of about 7~years and produce a radial velocity semi-amplitude of $5~{\rm m~s^{-1}}$. We further demonstrate that for a fiducial RV survey with $N=30$ epochs, measurement uncertainties of $\sigma = 4~{\rm m~s^{-1}}$, and a time baseline of $T=10~$years, the typical microlensing planet would then be marginally detectable with a signal-to-noise ratio (SNR) of 5. This suggests that there is at least some degree of overlap in the planet parameter space probed by RV and microlensing surveys.

In \citet{clanton_gaudi14a}, we then predict the joint probability distribution of RV observables for the whole planet population inferred from microlensing surveys. We find that the population has a median period of $P_{\rm med} \approx 9.4~$yr with a 68\% interval of $3.35\leq P/{\rm yr}\leq 23.7$ and a median RV semi-amplitude of $K_{\rm med}\approx 0.24~{\rm m~s^{-1}}$ with a 68\% interval of $0.0944\leq K/{\rm m~s^{-1}}\leq 1.33$. The California Planet Survey (CPS) includes a sample of 111 M dwarfs \citep{2014ApJ...781...28M} (hereafter MB14) which have been monitored for a median time baseline of over 10~years. The RV survey of HARPS includes 102 M dwarfs (BX13) that have been monitored for longer than 4~years. Thus, at least in terms of orbital period, these surveys should be sensitive to a significant fraction of the planet population inferred from microlensing. However, the fact that a majority of these planets produce radial velocities $K\lesssim 1~{\rm m~s^{-1}}$ means that many will remain undetectable by current generation RV surveys; this is primarily due to the steeply declining planetary mass function inferred by microlensing, $dN/d\log{q}\propto q^{-0.68\pm 0.20}$ \citep{2010ApJ...710.1641S}.

The results of \citet{clanton_gaudi14a} thus, qualitatively, indicate that the constraints on giant planet occurrence around M dwarfs inferred independently from microlensing and RV surveys are consistent. However, because the planetary mass function inferred by microlensing is so steep, the level of consistency is, quantitatively, very sensitive to the actual detection limits of a given RV survey. The primary aim of this paper is then to make an actual quantitative comparison of the planet detection results from microlensing and RVs. We start with a simulated population of microlensing-detected planets, the properties and occurrence rates of which are consistent with the actual population inferred from microlensing surveys for exoplanets \citep[GA10;][]{2010ApJ...710.1641S}, and map these into a population of analogous planets orbiting host stars monitored with RV. We next use the detection limits reported by BX13 for the HARPS M dwarf sample to predict the number of planets they should detect and compare this with the number of detections they report. We perform the same comparison with the CPS sample (MB14), but because they have yet to fully characterize the detection limits for each of their stars, this comparison is not as robust. For both comparisons, we also predict the number and magnitude of long-term RV trends that should be found and compare with the reported values. In doing so, we show that microlensing predicts that RV surveys should see a handful of giant planets around M dwarfs at the very longest periods to which they are sensitive. These planets have indeed been found. Because the detection results of these two discovery techniques are consistent, we are able to synthesize their independent constraints on the demographics of planets around M dwarfs to determine planet frequencies across a very wide region of parameter space, covering the mass interval $1<m_p\sin{i}/M_{\oplus}<10^4$ and period interval $1<P/{\rm days}<10^5$. We quote integrated planet frequencies over the period range $1<P/{\rm days}<10^4$ since our statistics are more robust in this interval.

Readers who are mainly interested in our results, but not necessarily the details, need only refer to figure~\ref{fig:freq_plot} and read the summary and discussion in \S~\ref{sec:discussion}. The full paper is organized as follows. We begin with a discussion of what exactly we mean by the term ``giant planet'' in \S~\ref{sec:giant_planet_def}. In \S~\ref{sec:sample_properties} we describe the sample properties of the microlensing and RV surveys we compare. We summarize the methodology developed in \citet{clanton_gaudi14a} to map the observable parameters of a planet detected by microlensing to the observable parameters of an analogous planet orbiting a star monitored with RV and describe the application of this methodology to this paper in \S~\ref{sec:methods}. We present our results, comparing our predicted numbers of detections and trends with the reported values of RV surveys in \S~\ref{sec:results}. \S~\ref{sec:uncertainties} details sources of uncertainty in our analysis. We derive combined constraints on the planet frequency around M dwarfs from RV and microlensing surveys in \S~\ref{sec:synthesis} and conclude with a discussion of our results in \S~\ref{sec:discussion}. Finally, we examine the properties of the planets accessible by both techniques in the Appendix.

\section{Definition of a ``Giant Planet''}
\label{sec:giant_planet_def}
At this point, it is worth discussing what we mean by a ``giant planet.'' This has not been precisely defined in the literature (to the best of our knowledge), but because microlensing surveys infer a steep planetary mass function, the precise definition is important. Giant planets, unlike terrestrial planets, should have significant hydrogen and helium atmospheres, and thus must form within the short timescales for gas dispersal in protoplanetary disks of $\sim 1-10~$Myr \citep[e.g.][]{1995Natur.373..494Z,2006ApJ...651.1177P}. Terrestrial planets and the cores of giant planets are believed to be formed via coagulation of planetesimals, initially tens of kilometers in size, growing through phases of both runaway and oligarchic growth \citep{safronov1969,1980ARA&A..18...77W,1985prpl.conf.1100H,1988Icar...74..542S,1989Icar...77..330W,1998Icar..131..171K}. Cores with masses of just $\sim 0.1~M_{\oplus}$ can attract gaseous envelopes, which are held up against gravity by pressure gradients maintained by the release of energy from planetesimals actively accreting onto the core. Further growth in core mass enables the attraction of still more nebular gas, such that the core accretion of planetesimals can no longer supply enough energy to support the increasingly massive envelope. The gaseous envelope contracts in response, increasing the rates of attraction of planetesimals and gas. Cores that reach a critical (or crossover) mass, such that the mass of the envelope is equal to the mass of the core ($M_{\rm env}\sim M_{\rm core}$), will accrete gas at a rate that increases exponentially with time, while the timescale for core accretion remains roughly constant. Various calculations have found that the critical mass should be somewhere in the range of $5-20~M_{\oplus}$, and is a function of the grain opacity and the rate of core accretion \citep{1980PThPh..64..544M,1986Icar...67..391B,1996Icar..124...62P,2000ApJ...537.1013I,2006ApJ...648..666R}.

Thus, a nascent planet with a core that reaches this critical mass before depletion of the nebular gas will ultimately be primarily composed of hydrogen and helium --- a giant planet. The final masses of giant planets then depends on the amount of gas they can accrete after this point, which is limited by available reservoir of gas that will eventually run out either because the planet opens a gap in the disk (assuming no gap-crossing accretion streams) or because the disk gas disperses before gap opening due to processes such as viscous dissipation, photoevaporation, and the like \citep[see e.g.][]{2007ApJ...667..557T}. The final masses of giant planets should then be upwards of some tens of Earth masses.

We define giant planets as having $>50\%$ hydrogen and helium by mass, which, in the core accretion paradigm, would imply that their cores must have reached the critical mass before the complete dispersal of disk gases. We choose to define a ``minimum'' giant planet mass of $0.1~M_{\rm Jup}\sim 30~M_{\oplus}$. We believe this to be a reasonable threshold because planets with $m_p\gtrsim 0.1~M_{\rm Jup}$ are likely composed of $>50\%$ hydrogen and helium by mass, unless their protoplanetary disk was very massive (and thus the isolation mass was large) or the heavy element content was $\gg 10\%$ \footnote{We note that there may exist counterexamples. For example, HD 149026b, originally discovered by \citet{2005ApJ...633..465S}, is believed to have a highly metal-enriched composition, probably $>50\%$ heavy elements by mass. HD 149026b has a mass of $m_p=0.37~M_{\rm Jup}\sim118~M_{\oplus}$, a radius of $R_p=0.8~R_{\rm Jup}$, and an orbital period of $P=2.9~$days \citep{2009ApJ...696..241C}. \citet{2009ApJ...696..241C} estimate this planet to have a core made up of elements heavier than hydrogen and helium with a mass in the range of $45-70~M_{\oplus}$, depending on the assumed stellar age and core density.}. For perspective, Jupiter and Saturn ($\sim 0.3~M_{\rm Jup}$) are primarily composed of hydrogen and helium, while Neptune ($\sim 0.05~M_{\rm Jup}$) and Uranus ($\sim 0.05~M_{\rm Jup}$) contain roughly 5-15\% hydrogen and helium, 25\% rocks, and 60-70\% ices, by mass, assuming the ice-to-rock ratio is protosolar \citep{1991uran.book...29P,1995P&SS...43.1517P,1995netr.conf..109H,2005AREPS..33..493G}.

\section{Microlensing and RV Sample Properties}
\label{sec:sample_properties}
\subsection{Microlensing Sample}
\label{subsec:gould_sample}
The microlensing sample of GA10 is an unbiased sample composed of 13 high-magnification events, fitting specific criteria that is described in detail in their paper. Unlike RV surveys, not much is known about the host (lens) stars in the microlensing sample. Nothing is known about the metallicity of the lens stars and there are estimates of, or upper limits on, the lens mass only for a subset of the sample. They report the lens stars (those with and without planets) to have a mass distribution centered around $0.5~M_{\odot}$ and thus adopt a typical lens mass for the sample of $M_l\sim 0.5~M_{\odot}$. As for the planet/host-star mass ratio and Einstein radius, they find typical values of $q\sim5\times10^{-4}$ and $R_E = 3.5$~AU$\left(M_{\star}/M_{\odot}\right)^{1/2}$, respectively. Using this sample, GA10 found the observed frequency of ice and gas giant planets (in the mass-ratio interval $-4.5 < \log{q} < -2$) around low-mass stars to be 
\begin{equation}
    \frac{d^2N_{\rm pl}}{d\log{q}~d\log{s}} = \left(0.36 \pm 0.15\right)~{\rm dex}^{-2}
\end{equation}
at the mean mass ratio $q_0 = 5\times 10^{-4}$ and sensitive to a wide range of projected separations, $s_{\rm max}^{-1}R_E \lesssim r_{\perp} \lesssim s_{\rm max}R_E$, where $R_E = 3.5~{\rm AU}~\left(M_{\star}/{\rm M_{\odot}}\right)^{1/2}$ and $s_{\rm max} \sim \left(q/10^{-4.3}\right)^{1/3}$, corresponding to deprojected separations of a few times larger than the position of the snow line in these systems. In order to better compare this frequency measurement with those from RV samples, we use the typical $M_l$ and $q$, along with the median value of $\sin{i}\approx 0.866$ and the median relation $a\sim r_{\perp}/0.866$ for randomly distributed orbits (see \citet{clanton_gaudi14a}), to estimate the frequency in terms of RV parameters,
\begin{equation}
    \frac{d^2N}{d\log{\left(m_p\sin{i}\right)}d\log{\left(a\right)}} = \left(0.36 \pm 0.15\right)~{\rm dex^{-2}}\; ,
\end{equation}
over the planetary mass interval $10\lesssim m_p\sin{i}/M_{\oplus}\lesssim 3\times10^3$ and the period interval $6\times10^2\lesssim P/{\rm days}\lesssim 6\times10^3$. Additionally, GA10 report no significant deviation from a flat distribution in $\log{s}$ for the events included in their analysis.

GA10 measure a normalization, but are unable to determine the slope of the planetary mass function. \citet{2010ApJ...710.1641S} assume a power-law form for the planetary mass-ratio function (also assuming planets follow a flat distribution in $\log{s}$) and measure the slope using the mass ratios of 10 microlensing-detected planets and their estimated detection efficiencies for each event, finding $dN/d\log{q}\propto q^{-0.68\pm 0.20}$. 

\citet{2012Natur.481..167C} use a few new microlensing-detected planets along with the previous constraints on the normalization by GA10 and the slope by \citet{2010ApJ...710.1641S} to measure the cool-planet mass function over an orbital range of $0.5-10~$AU, finding $d^2N/(d\log{m_p}\; d\log{a})=0.24^{+0.16}_{-0.10}\left(m_p/M_{\rm Sat}\right)^{-0.73\pm 0.17}$. In this paper, we choose to adopt the independent measurements of GA10 and \citet{2010ApJ...710.1641S} to construct our own planetary mass-ratio function, rather than adopt that of \citet{2012Natur.481..167C} (although, as we later show, the form we derive is consistent with that of \citet{2012Natur.481..167C}). We choose to do this because the measurements of GA10  and \cite{2010ApJ...710.1641S} are more closely related to the observable quantities we use as a starting point in this study.

\subsection{HARPS M Dwarf Sample}
\label{subsec:bonfils_sample}
The stellar sample of BX13 is a volume limited collection of 102 M dwarfs closer than 11~pc and brighter than $V = 14$~mag, with declinations $\delta < +20^{\circ}$ and with projected rotational velocities $v\sin{i} \lesssim 6.5$~m~s$^{-1}$. Known spectroscopic binaries and visual pairs with separations $< 5$'' were removed from the sample. The brightness range for this sample is $V = 7.3$~mag to 14~mag, with a median brightness of $V = 11.43~$mag. The stellar masses range between 0.09 to $0.6~M_{\odot}$, with a median mass of $0.27~M_{\odot}$. \citet{2013A&A...551A..36N} determine the metallicities of the stars in this sample, reporting [Fe/H] values ranging from -0.88~dex to 0.32~dex, with mean and median values of -0.13~dex and -0.11~dex, respectively. RV observations of this sample were made using the HARPS instrument \citep{2003Msngr.114...20M, 2004A&A...423..385P}. BX13 quote a precision of $\sigma \sim 80$~cm~s$^{-1}$ for $V = 7 - 10$ stars and $\sigma \sim 2.5^{\left(10 - V\right)/2}~{\rm m~s^{-1}}$ for $V = 10 - 14$ stars, which includes instrumental errors in addition to the photon noise. Their actual errors are larger, due to stellar jitter. 

BX13 report planet frequencies in several bins of $m_p\sin{i}$ and period. In order to better compare with the microlensing constraint, we have combined and transformed their detections into bins of $\log{\left(m_p\sin{i}\right)}$ and $\log{\left(a\right)}$, using the sample median stellar mass of $0.27~M_{\odot}$, to give a frequency
\begin{equation}
    \frac{d^2N}{d\log{\left(m_p\sin{i}\right)}d\log{\left(a\right)}} = \left(0.0057 \pm 0.0029\right)~{\rm dex^{-2}}\; ,
\end{equation}
for planets with $10 < m_p\sin{i}/M_{\oplus} < 10^4$ and $1 < P/{\rm days} < 10^3$. In the above calculation we did not include their period ranges of $10^3$--$10^4$ days, where the sensitivity of their survey rapidly declines. If we include the entire period range from $1 < P/{\rm days} < 10^4$, this frequency becomes
\begin{equation}
    \frac{d^2N}{d\log{\left(m_p\sin{i}\right)}d\log{\left(a\right)}} = \left(0.0088 \pm 0.0039\right)~{\rm dex^{-2}}\; .
\end{equation}

\subsection{CPS M Dwarf Sample}
\label{subsec:johnson_sample}
The stellar sample of the RV study conducted by JJ10 included about 120 M dwarfs brighter than $V=11.5$ monitored by the CPS team with HIRES \citep{1994SPIE.2198..362V} at Keck Observatory, and are reported to have masses between $M_{\star} < 0.6~M_{\odot}$ and a wide range of metallicites between $-0.6 < {\rm \left[Fe/H\right]} < 0.6$. Their analysis consisted of planets with semi-major axes $a < 2.5$~AU and systems with velocity semi-amplitudes of $K > 20~{\rm m~s^{-1}}$. Using this sample, JJ10 found the observed frequency of giant planets around low-mass stars, corrected for the average stellar metallicity, to be $2.5 \pm 1.2\%$. For comparison with the microlensing results, we convert this into units of dex$^{-2}$ by dividing by the area it covers in the $\log{\left(m_p\sin{i}\right)}$--$\log{\left(a\right)}$ plane. This non-rectangular area is bound by the above mentioned constraints, imposed by the set of planets included in the analysis. This yields a frequency of
\begin{equation}
    \frac{d^2N_{\rm pl}}{d\log{\left(m_p\sin{i}\right)}~d\log{\left(a\right)}} = \left(0.0085 \pm 0.0041\right)~{\rm dex}^{-2}\; ,\label{eqn:jj_freq_estimate}
\end{equation}
for masses $m_p\sin{i}\gtrsim150~M_{\oplus} \left(M/0.5~{M_{\odot}}\right)^{1/2}\left(a/{\rm AU}\right)^{1/2}$ and periods $P \lesssim 2\times10^3~{\rm days}\left(M/0.5~M_{\odot}\right)^{-1/2}$, where we have chosen a characteristic host mass of $M\sim 0.5~M_{\odot}$ to transform to similar parameters as the RV survey of BX13 and the microlensing survey of GA10.

The CPS continues to monitor these stars and, since the study of JJ10, has extended their sample to M dwarfs (which they define as having $B-V>1.44$) brighter than $V=13.5$, bringing their M dwarf sample to a total of 131 stars with no known stellar companions within two arcseconds and all closer than 16~pc. MB14 further refine this sample by excluding stars with known, nearby stellar binary companions. The final sample, which we will refer to as the ``CPS sample'' throughout this paper, consists of 111 M dwarfs with a median time baseline of $11.8~$yr, a median of $29$ epochs per star, and typical Doppler precisions of a couple meters per second. They also estimate $\sim 3-6~{\rm m~s^{-1}}$ of stellar jitter for the majority of their stars. In this study, we will compare the numbers of detections and trends the CPS have discovered from this M dwarf sample (of which the sample of JJ10 is a subset) to the amount we predict they should find based on the microlensing measurements of planet frequency around low mass stars.

\section{Methods}
\label{sec:methods}
In \citet{clanton_gaudi14a}, we developed the methodology to map the observable parameters of a planet detected by microlensing to the observable parameters of an analogous planet orbiting a star monitored with RV, i.e. $\left(q,s\right)\rightarrow \left(K,P\right)$, where $K$ and $P$ are the velocity semi-amplitude and orbital period, respectively. We then used this procedure to show that a fiducial RV survey with a precision of $\sigma=4~{\rm m~s^{-1}}$, an average number of epochs per star of $N=30$, a duration of $T=10~$years, and monitoring 100 stars uniformly (in log space) covering the mass interval $0.07~M_{\odot}\leq M_{\star}\leq 1.0~M_{\odot}$, should on average detect $4.9^{+4.6}_{-2.6}$ planets and identify $2.4^{+2.4}_{-1.4}$ long-term RV trends resulting from planets at a SNR of at least 5, motivating a more rigorous comparison to actual RV surveys.

In this section, we first provide a brief account of the methods and results presented in \citet{clanton_gaudi14a}, followed by a description of how we apply this methodology to directly compare planet detection results from the microlensing survey of GA10 to those from the HARPS (BX13) and CPS (MB14) RV surveys of M dwarfs. For more details on the methodology, refer to \citet{clanton_gaudi14a}.

\subsection{Mapping Analogs of Planets Found by Microlensing into RV Observables}
\label{subsec:mapping_summary}
The general procedure detailed in \citet{clanton_gaudi14a} is comprised of a two steps. The first step is the mapping $\left(q,s\right) \rightarrow \left(m_p,r_{\perp}\right)$ using a Galactic model. Here, $q$ and $s$ are the planet-to-star mass ratio and the planet-star projected separation in units of the Einstein radius ($\theta_E$), respectively, and are the quantities measured in a microlensing planet detection. The mapping between these measurements and the true planet mass, $m_p=qM_l$, and the projected separation in physical units, $r_{\perp}=sD_l\theta_E$, requires a Galactic model because the precise forms of the distributions of physical parameters of microlensing systems are unknown. In particular, we do not know the true distribution of lens masses, $M_l$, or distances, $D_l$, nor do we know with certainty whether the lens lies in the disk or the bulge in a given microlensing event. We account for this by drawing these parameters from basic priors and weighting by the corresponding microlensing event rate, $\Gamma$, assuming a Galactic model.

The second step is the mapping $\left(m_p,r_{\perp}\right)\rightarrow \left(K,P\right)$, where $K$ and $P$ are the velocity semi-amplitude and the orbital period, respectively. This is accomplished by adopting priors on, and marginalizing over, the Keplerian orbital parameters (i.e. inclination, eccentricity, mean anomaly, and argument of periastron) of the microlensing-detected systems to get a distribution of semimajor axes, which then immediately gives the $P$ distribution by way of Kepler's third law. Combining the period distribution with $m_p$ and the distribution of inclinations, we are able to derive the distribution of $K$.

\begin{figure*}
\epsscale{0.9}
\plotone{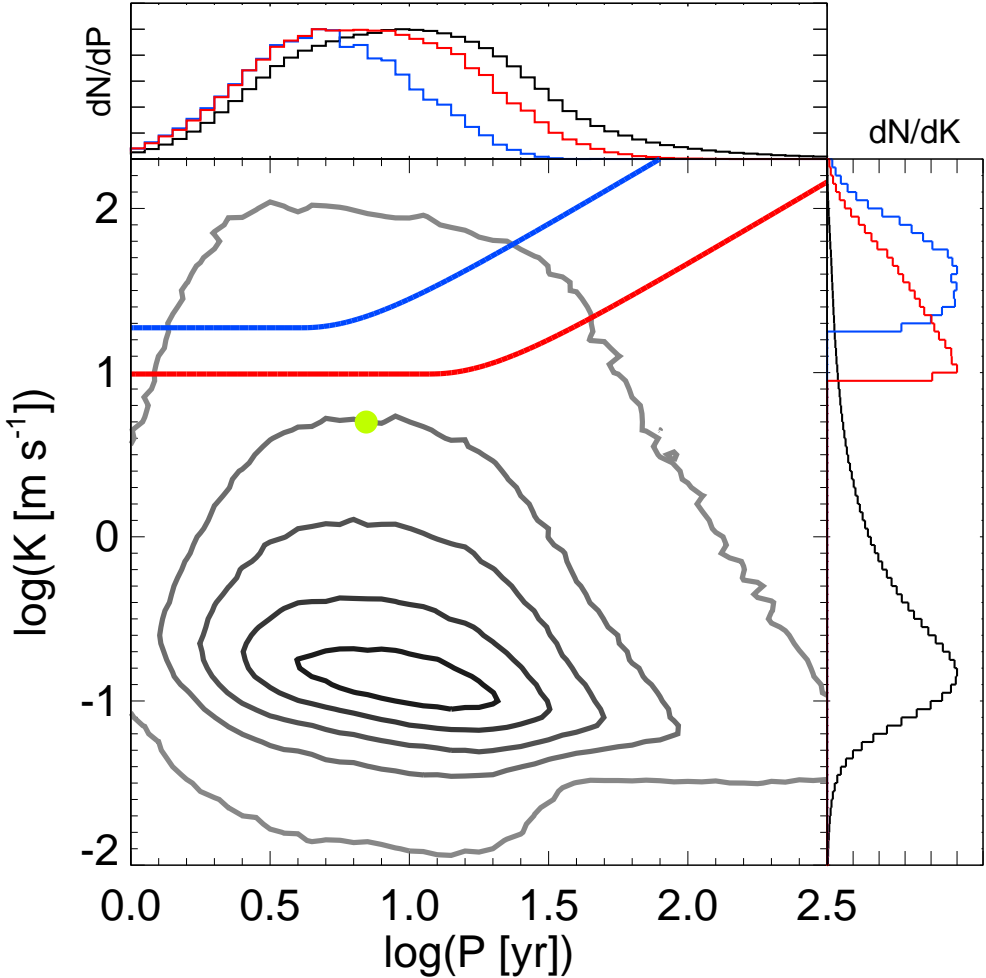}
\caption{Mapping of microlensing planets into RV observables, from \citet{clanton_gaudi14a}. Shown in greyscale are contours of the probability density of $K$ and $P$, marginalized over the entire microlensing parameter space. The contour levels, going from grey to black, are $1\%$, $10\%$, $25\%$, $50\%$ and $80\%$ of the peak density. The filled yellow circle represents where the typical microlensing planet lies in this parameter space at the median inclination and mean anomaly and on a circular orbit ($K_{\rm typ}\sim 5~{\rm m~s^{-1}}$, $P_{\rm typ}\sim 7~$yr). The blue and red colored line represent the median RV detection limit curves for the surveys of BX13 and JJ10, respectively. Planets that lie above these lines and have periods less than the duration of the RV survey are detectable, while those with longer periods might show up as long-term RV trends. The colored histograms represent the the total numbers of detections plus trends for the HARPS sample (blue curve) and the CPS sample (red curve) as a function of $P$ (top panel) and $K$ (right panel). It is clear from these colored histograms that RV surveys are beginning to sample the full period distribution of the planet population inferred from microlensing, but are only able to catch the tail of the $K$ distribution towards higher values, or equivalently, the high-mass end of this planet population.
  \label{fig:marg_dist}}
\end{figure*}

Figure \ref{fig:marg_dist} shows the resultant joint distribution of $K$ and $P$ for a population of planets analogous to that inferred from microlensing, marginalized over all planet and host star properties inferred from microlensing, as well as all orbital parameters \citep{clanton_gaudi14a}. The median values we found are $P_{\rm med} \approx 9.4~$yr and $K_{\rm med}\approx 0.24~{\rm m~s^{-1}}$. The 68\% intervals in $P$ and $K$ are $3.35\leq P/{\rm yr}\leq 23.7$ and $0.0944\leq K/{\rm m~s^{-1}}\leq 1.33$, respectively, and their 95\% intervals are $1.50\leq P/{\rm yr}\leq 94.4$ and $0.0422\leq K/{\rm m~s^{-1}}\leq 16.8$, respectively. In \citet{clanton_gaudi14a}, we demonstrated how to compute the expected number of planets an RV survey should detect, as well as the number of long-term RV trends (due to planets) that should be seen, by parameterizing RV detection limits in terms of a SNR threshold.

We showed that the phase-averaged SNR, which we designate as $\mathcal{Q}$, assuming uniform and continuous sampling of the RV curve, is
\begin{align}
    \mathcal{Q} = & {} \left(\frac{N}{2}\right)^{1/2}\left(\frac{K}{\sigma}\right) \nonumber \\
    & {} \times \left\{1-\frac{1}{\pi^2}\left(\frac{P}{T}\right)^2\sin^2{\left(\frac{\pi T}{P}\right)}\right\}^{1/2}\; , \label{eqn:snr_full}
\end{align}
where $N$ is the average number of epochs per star, $\sigma$ is the average RV precision and $T$ is the time baseline of the RV survey. In the limit where the period is much less than the time baseline, $P\ll T$, this reduces to 
\begin{equation}
    \mathcal{Q} \approx \left(N/2\right)^{1/2}\left(K/\sigma\right)\; , \label{eqn:snr_small_p}
\end{equation}
which is also a good approximation for periods up to $P\sim T$ when approaching from $T$ from small $P$. We assume an effective sensitivity for our fiducial RV survey by assuming a SNR threshold, $\mathcal{Q}_{\rm min}$, above which planets can be detected. Solving equation (\ref{eqn:snr_full}) for $K$ in terms of $P$, we find a sensitivity of
\begin{align}
    K_{\rm min} = & {} \mathcal{Q}_{\rm min}\sigma\left(\frac{2}{N}\right)^{1/2} \nonumber \\
    & {} \times \left\{1-\frac{1}{\pi^2}\left(\frac{P}{T}\right)^2\sin^2{\left(\frac{\pi T}{P}\right)}\right\}^{-1/2} \; , \label{eqn:k_sens_first}
\end{align}
meaning that the RV survey will be sensitive to planets that produce velocity semi-amplitudes greater than or equal to $K_{\rm min}$ at SNRs of $\mathcal{Q}_{\rm min}$ or greater. We further make the approximation that only planets with periods $P\leq T$ will be detected, whereas planets with periods $P>T$ can possibly be identified as long-term RV trends.

In this study, to compute the expected number of detections and long-term trends for the RV survey of BX13, we approximate the detection limit curves they provide for each star in their sample by fitting equation (\ref{eqn:k_sens_first}) to their curves with $\mathcal{Q}_{\rm min}$ as a free parameter. We provide more information on our approximation of the detection limits of both the HARPS and CPS samples in \S~\ref{subsec:method_application} and \S~\ref{subsubsec:bonfils_detailed_comparison}.

Figure \ref{fig:marg_dist} shows a couple examples of such a sensitivity curve, given by equation (\ref{eqn:k_sens_first}), over-plotted on top of our joint distribution of $K$ and $P$. The blue curve represents the median detection limit as a function of period for the HARPS sample (BX13), which has the median values $N_{\rm med}=8$, $\sigma_{\rm med}\approx 4.2~{\rm m~s^{-1}}$, $T_{\rm med}\approx 4.1~$yr, $M_{\star, {\rm med}}=0.27~M_{\odot}$, and $\mathcal{Q}_{\rm min, med}\approx 8.9$, and the red curve is that of the CPS sample (MB14), which has the median values $N_{\rm med}=28$, $\sigma_{\rm med}\approx 4.1~{\rm m~s^{-1}}$, $T_{\rm med}\approx 11.1~$yr, $M_{\star, {\rm med}}=0.43~M_{\odot}$, and $\mathcal{Q}_{\rm min, med}\approx 8.3$.

We can rewrite equation (\ref{eqn:k_sens_first}) in terms of a minimum $m_p\sin{i}$ by substituting the velocity semi-amplitude equation for $K$ and solving, to yield an equivalent sensitivity in terms of planetary mass
\begin{align}
    \left.m_p\sin{i}\right|_{\rm min} = & {} \; \mathcal{Q}_{\rm min}\sigma M_{\star}^{2/3}\left(\frac{2}{N}\right)^{1/2}\left(\frac{P}{2\pi G}\right)^{1/3} \nonumber \\
    & {} \times \left\{1-\frac{1}{\pi^2}\left(\frac{P}{T}\right)^2\sin^2{\left(\frac{\pi T}{P}\right)}\right\}^{-1/2} \label{eqn:mpsini_sens_first}
\end{align}
which evaluates to
\begin{align}
    \left.m_p\sin{i}\right|_{\rm min} \approx & {}\;  69~M_{\oplus}\left(\frac{P}{\rm 7~yr}\right)^{1/3}\left(\frac{M_{\star}}{0.5~M_{\odot}}\right)^{2/3} \nonumber \\
    & {} \times \left(\frac{\mathcal{Q}_{\rm min}}{5}\right)\left(\frac{\sigma}{4~{\rm m~s^{-1}}}\right) \nonumber \\
    & {} \times \left(\frac{N}{30}\right)^{-1/2}
\end{align}
in the approximation $P\ll T$.

Also plotted in the top and right panels of figure~\ref{fig:marg_dist} are colored histograms representing the total numbers of detections plus trends for the HARPS sample (blue curve) and the CPS sample (red curve) as a function of $P$ (top panel) and $K$ (right panel). It is clear from these colored histograms that RV surveys are beginning to sample the full period distribution of the planet population inferred from microlensing, but are only able to catch the tail of the $K$ distribution towards higher values, or equivalently, the high-mass end of this planet population.

\subsection{Application: Comparing with Real RV Surveys}
\label{subsec:method_application}
The application of this methodology to compare microlensing detections to those reported by real RV surveys is a little more involved than our description above. In that simple estimate, we assumed each star had the same number of epochs, the same measurement uncertainties at each epoch, and that each star was observed over the same time baseline. The reality is that RV surveys have varying sensitivities for each of their monitored stars which need to be included in a direct comparison. We must also take care to construct a microlensing sample that is consistent with that of real RV surveys, i.e. one with the same distribution of host star masses. In this section, we describe how we do this in order to perform independent statistical comparisons of planet detection results from microlensing with each of the RV surveys of HARPS and CPS.

When comparing with the HARPS survey, we begin with an ensemble of microlensing events for a sample of planet-hosting stars in the mass interval $0.07\leq M_l/M_{\odot}\leq 1.0$ for which we have numerically determined the joint distributions of the RV observables $K$ and $P$. In order to force the microlensing sample to be consistent with that of HARPS, we consider only microlensing detections around lenses with $\left|M_l-M_{\star}\right|\leq \sigma_{M_{\star}}$ for each star in the RV sample, where $M_l$ is the lens mass for a given microlensing event, $M_{\star}$ is the mass of the RV monitored star, and $\sigma_{M_{\star}}$ is the uncertainty on the measurement of $M_{\star}$. This yields a set of distributions of $K$ and $P$, each corresponding to a particular microlensing planet detection that has been mapped into these observables. We then sum up all the joint $K$ and $P$ distributions for each set of events with lens star masses within $\pm \sigma_{M_{\star}}$ of $M_{\star}$. The summation and weighting of these distributions is done in exactly same manner as described in \S~\ref{subsec:mapping_summary} (and in more detail in \citet{clanton_gaudi14a}), except that now, rather than marginalizing over the entire mass interval $0.07\leq M_l/M_{\odot}\leq 1.0$, we have instead marginalized over all lens masses within $\pm \sigma_{M_{\star}}$. We are left with a single distribution, $d^2N_{\rm pl}/(dKdP)$, for each star in the RV sample. We note that by matching the host mass distribution of our simulated sample to that of HARPS, we are implicitly assuming that the microlensing planet distribution is independent of host mass, $M_{\star}$. This is unavoidable because the microlensing sample is not large enough to subdivide and determine the planet frequency dependence on host mass.

\begin{figure}[t!]
\epsscale{1.2}
\plotone{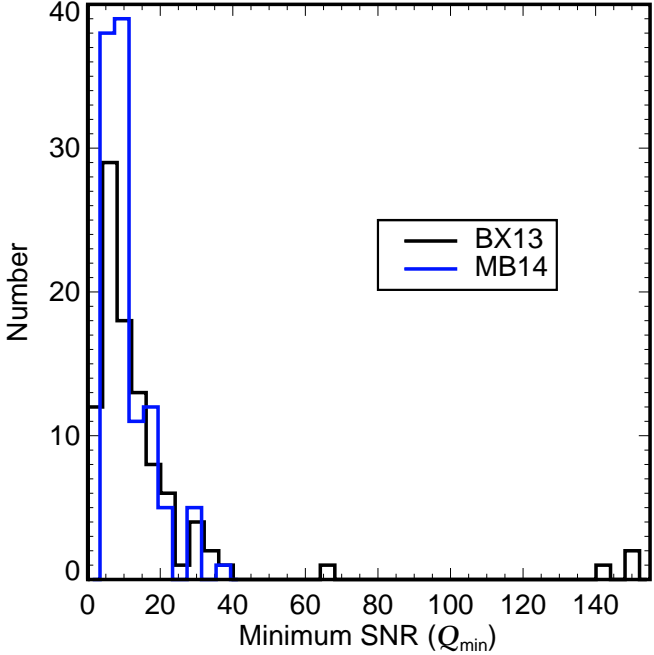}
\caption{Distribution of SNR thresholds ($\mathcal{Q}_{\rm min}$) we find for the HARPS and CPS M dwarf samples. The median values for these surveys are 8.9 and 8.3, respectively. These values represent the minimum SNRs to which a given RV survey can detect a planet or identify a long-term RV trend, and are used to approximate the detection sensitivities of these two RV surveys for each star in their samples.
  \label{fig:sn_fig}}
\end{figure}

In order to compute the expected number of detections and trends for each star in the HARPS RV sample, we must first model the sensitivity of their survey for each star, in terms of $K$ and $P$. For each star in their sample, BX13 graphically provide detection limit curves, i.e. the minimum $m_p\sin{i}$ to which they are sensitive as a function of $P$. They generate these detection limits by systematically injecting known (fictitious) planetary signals into their data and determining the subset of these signals that are detectable (see \S~6 of BX13 for a more detailed explanation). We approximately reproduce these detection limits by parameterizing in terms of a minimum SNR. We use the values of $\sigma$, $M_{\star}$, $T$, and $N$ for each star provided by BX13, including $\mathcal{Q}_{\rm min}$ as a free parameter, to match (by eye) equation (\ref{eqn:mpsini_sens_first}) to the detection limit curves for each star. We describe the RV measurement uncertainties we adopt in \S~\ref{subsubsec:bonfils_detailed_comparison}. Many of these curves are quite noisy (see figure 18 of BX13), so we match to the approximate mean of the noise in these curves by eye. This parameterization of their detection limits can be interpreted as computing the minimum SNR to which the survey can detect a planet or identify a long-term RV trend. The distribution of $\mathcal{Q}_{\rm min}$ we find for the HARPS sample is shown in figure \ref{fig:sn_fig}. The fact that $\mathcal{Q}_{\rm min}$ varies from star to star is a reflection of the non-uniformity of the HARPS M dwarf sample, i.e. each star has a different number of epochs, and spans a different time baseline, resulting in differing detection limits within the sample. The four stars with $\mathcal{Q}_{\rm min}\geq 50$ shown in figure~\ref{fig:sn_fig} are from stars with just four epochs that span relatively short time baselines.

These SNR values are used in conjunction with equation (\ref{eqn:k_sens_first}) to compute the number of detections and trends we expect the HARPS M dwarf survey to find in the same manner as described in \S~{\ref{subsec:mapping_summary}} and illustrated in figure \ref{fig:marg_dist}. These expected numbers of detections and trends are then compared with the actual numbers reported by BX13. The results and comparison is presented in \S~\ref{subsec:bonfils_comparison}.

We follow an identical procedure for computing the expected numbers of detections and trends for the CPS survey, except for the way in which we estimate their detection limits. The CPS team has not yet determined the individual detection sensitivities for their sample, so to roughly estimate their detection limits (in terms of $\mathcal{Q}_{\rm min}$) we assume the sensitivities of their stars are similar to those of stars with similar systematics in the HARPS sample. We compute values of $\sigma_i/N^{1/2}$ for all stars in both RV samples, where $\sigma_i$ is the RV measurement precision (not including ``external'' noise sources, e.g. stellar jitter) and $N$ is the number of epochs. Each star in the CPS sample is ``matched'' to the star in the HARPS sample with the nearest value of $\sigma_i/N^{1/2}$. We assume the matched pairs of stars have similar sensitivities, and assign the stars in the CPS sample the same sensitivities (i.e. the same minimum SNR, $\mathcal{Q}_{\rm min}$) as that of the star in the HARPS sample to which they are matched. Since the CPS team reports stellar jitter values of $3-6~{\rm m~s^{-1}}$ for all stars in their sample, we only ``match'' them to stars in the HARPS sample which have consistent ``external'' errors of $\sigma_e\leq6~{\rm m~s^{-1}}$. The resultant distribution of $\mathcal{Q}_{\rm min}$ we obtain for the CPS sample is displayed against that of the BX13 in figure \ref{fig:sn_fig}, and has a median value of 8.3. The expected number of planet detections and long-term RV trends is calculated in the same manner as those for the HARPS survey. Our results and comparison with the CPS sample is presented in \S~\ref{subsec:johnson_comparison}. Ideally, we would like to do this comparison more accurately once the CPS determines their detection limits for their sample.

In \citet{clanton_gaudi14a}, we derive the planetary mass-ratio and projected separation function
\begin{align}
    \frac{d^2N_{\rm pl}}{d\log{s}~d\log{q}} = & {} \left(0.23\pm 0.10\right)~{\rm dex^{-2}} \nonumber \\
    & {} \times \left(\frac{q}{q_0}\right)^{-0.68\pm 0.20} \; , \label{eqn:planetary_mass_function}
\end{align}
where $q_0=5\times10^{-4}$. We adopt the slope of the planetary mass-ratio function $dN_{\rm pl}/d\log{q} \propto q^{p}$, where $p=-0.68\pm 0.20$, from \citet{2010ApJ...710.1641S} and normalize it using the integrated frequency measurement of $d^2N_{\rm pl}/(d\log{q}~d\log{s})\equiv \mathcal{G}=(0.36\pm0.15)~{\rm dex}^{-2}$ by GA10. We assume planets are uniformly distributed in $\log{s}$ since the distribution of projected separations from the sample of GA10 is consistent with such a distribution. As we will show in \S~\ref{sec:uncertainties}, the main uncertainties in our results arise from the uncertainties in $p$ and $\mathcal{G}$.

Mathematically, the total number of planet detections we expect a RV sample to yield for a given realization $i$ in our simulation (corresponding to given values of $p_i$ and $\mathcal{G}_i$) is 
\begin{equation}
    N_{\rm det, i}=\displaystyle \sum_k N_{\rm det, i, k}\; ,
\end{equation}
where $N_{\rm det, i, k}$ is the number of expected planet detections for a given star $k$,
\begin{align}
    N_{\rm det, i, k} = & {} \displaystyle \int dM_l \int dD_l \int d\log{q} \int d\log{s} \nonumber \\
    & {} \times \int dK \int dP \left.\frac{d^6N_{\rm pl}}{dKdPdM_ldD_ld\log{q}~d\log{s}}\right|_{i} \nonumber \\
    & {} \times \Phi_{{\rm det}, k}\left(\mathcal{Q}\right)\Phi_{{\rm det}, k}\left(P\right)\Phi_k\left(M_l\right)\; , \label{eqn:n_det_analytic}
\end{align}
where $\Phi_{{\rm det},k}\left(\mathcal{Q}\right)$ and $\Phi_{{\rm det},k}\left(P\right)$ are selection functions on a given star constraining the detections to those planets which have SNRs larger than the threshold value (i.e. $\mathcal{Q}_{\rm min}$) and periods smaller than the time baseline of observations, $T$, for that particular star in the RV sample with which we are comparing. The functional forms of these are $\Phi_{{\rm det},k}\left(\mathcal{Q}\right)=\Theta\left(\mathcal{Q} - \mathcal{Q}_{{\rm min},k}\right)$ and $\Phi_{{\rm det},k}\left(P\right)=\Theta\left(T_k-P\right)$, respectively, where $\Theta$ is the Heaviside step function. In equation (\ref{eqn:n_det_analytic}), $\Phi_k\left(M_l\right)$ is the selection function on lens masses that we employ to force our microlensing sample to have the same stellar mass distribution as the RV survey to which we are comparing, having the functional form $\Phi_k\left(M_l\right)=\Theta\left[M_l - \left(M_{\star, k}-\sigma_{M_{\star, k}}\right)\right]\Theta\left[\left(M_{\star, k}+\sigma_{M_{\star, k}}\right)-M_l\right]$.

The integrand of equation (\ref{eqn:n_det_analytic}) (not including the selection functions) represents the distribution of $K$ and $P$ for a single system, i.e. only one $M_l$, $D_l$, $\log{q}$, and $\log{s}$, marginalized over all possible orbital configurations. Integrating this distribution marginalizes over all planet and host star properties inferred from microlensing. Multiplying this distribution by selection functions of RV detectability and on the host star mass, as in equation (\ref{eqn:n_det_analytic}), and integrating yields the number of RV detectable planets for a given host star mass. As we showed in \citet{clanton_gaudi14a}, the distribution function is given formally as
\begin{align}
    & {} \left.\frac{d^6N_{\rm pl}}{dKdPdM_ldD_ld\log{q}d\log{s}}\right|_{i} = \mathcal{F}_i\displaystyle \int_{\left\{\alpha\right\}}d\left\{\alpha\right\} \nonumber \\
    & {} \hspace{0.8in} \times \frac{d^n{\rm N_{pl}}}{d\left\{\alpha\right\}}\delta\left(K\left(m_p, i, M_l, a\right)-K'\right)\nonumber \\
    & {} \hspace{0.8in} \times \delta\left(P\left(M_l, m_p, a\right)-P'\right)\delta\left(M_l-M_l'\right) \nonumber \\
    & {} \hspace{0.8in} \times \delta\left(D_l-D_l'\right)\delta\left(q-q'\right)\delta\left(s-s'\right)\; ,\label{eqn:planet_dist_function}
\end{align}
where $\left\{\alpha\right\}$ is the set of all $n$ intrinsic, physical parameters on which the frequency of planets fundamentally depends. We assume the form
\begin{align}
    \frac{d^n{\rm N_{pl}}}{d\left\{\alpha\right\}} = & {} \frac{d{\rm N_{pl}}}{di}\frac{d{\rm N_{pl}}}{da}\frac{d{\rm N_{pl}}}{dM_0}\frac{d^2{\rm N_{pl}}}{d\log{q}~d\log{s}} \nonumber \\
    & {} \times \frac{d{\rm N_{pl}}}{dM_l}\frac{d{\rm N_{pl}}}{dD_l}\frac{d{\rm N_{pl}}}{d\omega}\frac{d{\rm N_{pl}}}{de}\; , \label{eqn:orb_marg_function}
\end{align}
and we note that
\begin{equation}
    \frac{dN_{\rm pl}}{dM_l}\frac{dN_{\rm pl}}{dD_l} \propto \displaystyle \int \int \frac{d^4d\Gamma}{dD_ldM_ld^2\boldsymbol{\mu}}\Phi\left(t_E\right)d^2\boldsymbol{\mu}\; ,
\end{equation}
where $d^4d\Gamma/(dD_ldM_ld^2\boldsymbol{\mu})$ is the event rate of a given microlensing event, $\Phi\left(t_E\right)=\Theta\left(t_E/{\rm days} - 10\right)$ is a selection function on the event timescale, $t_E$, and $\boldsymbol{\mu}$ is the lens-source relative proper motion. Finally, the $\mathcal{F}_i$ in equation (\ref{eqn:planet_dist_function}) represents the effective number of planets per star in the area over which our simulated planetary microlensing evetns are sampled, i.e., the integral over that area weighted by the joint distribution function $d^2N_{\rm pl}/(d\log{q}~d\log{s})$, 
\begin{equation}
    \mathcal{F}_i = \mathcal{A}_i\displaystyle \int_{\log{0.5}}^{\log{2.5}} \int_{-5}^{-2}\left(\frac{q}{q_0}\right)^{p_i}d\log{q}~d\log{s}\; .\label{eqn:f_a_n}
\end{equation}
We find a mean value and 68\% confidence interval of $\mathcal{F}=1.5\pm 0.6$. For our final results, we adopt the mean value of the number of detections from all realizations (i.e the expectation value) and the 68\% confidence intervals to represent our errors. Uncertainties in $p$ and $\mathcal{G}$ are numerically propagated through our simulations and are responsible for the uncertainties in our final results.

Similarly, the total number of expected long-term RV trends per star for an RV survey is given by equation (\ref{eqn:n_det_analytic}), but with the new selection function $\Phi_{{\rm det}, k}\left(P\right)\rightarrow\Phi_{{\rm tr}, k}\left(P\right)=\Theta\left(P-T_k\right)$, such that only planets with periods larger than the time baseline of observations for a given star are counted as trends. Refer to \citet{clanton_gaudi14a} for a more complete description of the mathematical formalism presented here.

\section{Results}
\label{sec:results}
We compare the numbers of planet detections and long-term trends reported for the HARPS (BX13) and CPS (MB14) M dwarf surveys to the amount we predict they should find by assuming a population of planets analogous to that inferred from microlensing surveys. Since BX13 provide detection limits for each of the stars in the HARPS M dwarf sample, we primarily focus on the comparison with their survey, first performing an order of magnitude comparison for the number of predicted planet detections before doing a more detailed analysis. We then compare with the CPS sample by assuming their detection sensitivities are similar to that of BX13 for stars with similar RV uncertainties between the two surveys, as described above.

\subsection{Comparison with HARPS Planet Detections}
\label{subsec:bonfils_comparison}
\subsubsection{Order of Magnitude Comparison}
\label{subsubsec:bonfils_OoM_comparison}
In order to better understand the result of our detailed calculation, we first derive an order of magnitude estimate of the number of RV-detectable planets in the HARPS sample by assuming their survey is uniformly sensitive to planets over a given range of mass ratios and projected separations. We then estimate the planet frequency at the median mass ratio and projected separation in this range, which we designate as
\begin{equation}
    f_{\rm med}=\left.\frac{d^2N_{\rm pl}}{d\log{q}d\log{s}}\right|_{q=q_{\rm med}, s=s_{\rm med}}\; ,
\end{equation}
and make the approximation that this does not change over the entire parameter space to which BX13 is sensitive. Multiplying this by the sample size of HARPS and the area in $\log{q}-\log{s}$ space over which we assume they are sensitive yields a rough estimate of the number of expected planet detections
\begin{align}
    N_{\rm pl} \sim & {} \; N_{\star}\left(\log{q_{\rm max}}-\log{q_{\rm min}}\right) \nonumber \\
    & {} \; \times \left(\log{s_{\rm max}}-\log{s_{\rm min}}\right)f_{\rm med}\; . \label{eqn:n_det_estimate}
\end{align}

We assume BX13 is sensitive to the higher end of the range of mass ratios to which microlensing is sensitive, so that $q_{\rm max}=10^{-2}$. To estimate $q_{\rm min}$, we roughly compute their average sensitivity limit by using representative values of $M_{\star}$, $N$, $\sigma$, and the median $\mathcal{Q}_{\rm min}$ (see \S~\ref{subsubsec:bonfils_detailed_comparison} and figure \ref{fig:sn_fig}). Substituting for $K$ in equation (\ref{eqn:snr_small_p}) using the standard velocity semi-amplitude equation for a circular orbit, solving for $m_p\sin{i}$ and dividing both sides by $M_{\star}$, we obtain an expression for the minimum mass ratio, to which an RV survey will be sensitive (in the limit $P\ll T$),
\begin{equation}
    q_{\rm min} \sim \left(\frac{P_{\rm typ}}{2\pi G}\right)^{1/3}M_{\star}^{-1/3}\mathcal{Q}_{\rm min}\sigma\sqrt{\frac{2}{N}}\; , \label{eqn:q_sensitivity_bonfils}
\end{equation}
where $P_{\rm typ}\approx 7~$yr is the period for the typical microlensing planet found in \citet{clanton_gaudi14a}. Using the median values reported by BX13 for the HARPS sample of $N_{\rm med}=8$, $\sigma_{\rm med}=4.2~{\rm m~s^{-1}}$, $M_{\rm \star, med}=0.27~M_{\odot}$, and ${\rm \mathcal{Q}_{\rm min, med}}\sim 10$, we estimate the ``average'' minimum mass ratio to which they are sensitive to be $\log{q_{\rm min}}\approx -2.7$.

We then assume that BX13 can efficiently detect planets at the lower end of the range of projected separations to which microlensing is also sensitive, which sets $s_{\rm min}=0.5$. We approximate the largest projected separation to which BX13 are sensitive as $s_{\rm max}\sim \left(T_{\rm med}/P_{\rm typ}\right)^{2/3}\approx 0.69$, where $P_{\rm typ}\approx 7~$yr is the period of the typical microlensing planet and $T_{\rm med}=4.1~$years is the median time baseline for the HARPS M dwarfs. For a $0.5~M_{\odot}$ star, these ranges roughly correspond to planet masses between $\sim 1-5~M_{\rm Jup}$ and projected separations between $\sim 1-2~$AU (for $D_l/D_s = 1/2$). The median log values are then $\log{q_{\rm med}} \approx \left(\log{q_{\rm max}}+\log{q_{\rm min}}\right)/2\approx -2.35$ and $\log{s_{\rm med}} \approx \left(\log{s_{\rm max}}+\log{s_{\rm min}}\right)/2\approx -0.2$. We find a mean and 68\% confidence interval of $f_{\rm med}= 0.064^{+0.042}_{-0.043}$ using equation (\ref{eqn:planetary_mass_function}) and these median values.

Using these values and equation (\ref{eqn:n_det_estimate}), we expect BX13 to detect $N_{\rm pl} = 0.63^{+0.41}_{-0.42}$ planets from the $N_{\star}\sim 100$ stars they monitor, where the errors on this estimate come from uncertainties in the normalization (GA10) and exponent \citep{2010ApJ...710.1641S} of the planetary mass-ratio function given by equation (\ref{eqn:planetary_mass_function}). This answer is within a factor of $\sim 2$ of the result we obtain from the detailed calculation in the next section.

\subsubsection{Detailed Comparison}
\label{subsubsec:bonfils_detailed_comparison}
BX13 monitor a total of 102 stars. We discard the four stars with less than four epochs. We also eliminate Gl 803 from the sample. The mass they report for this star is $0.75~M_{\odot}$, which is derived from the empirical mass-luminosity relationship of \citet{2000A&A...364..217D} in conjunction with parallax information and K-band photometry. They note in a footnote below their Table 3 that Gl 803 (AU Mic) is a $\sim 20~$Myr star with a circumstellar disk and so the calibration for determining its mass may not be valid given its age. To keep their mass estimations consistent, they chose not to adopt the mass found by \citet{2004Sci...303.1990K} for this star of $0.5~M_{\odot}$. We argue that Gl 803 should not be included in their sample on the grounds that it is not an M dwarf given the mass estimate they choose to adopt. We note that there are no known planets around this star, although it does show variation of a couple hundred meters per second. However, with only four epochs, we cannot say anything about the source of this variation. Thus, the refined HARPS M dwarf sample we consider includes 97 M dwarfs with four or more epochs.

We obtain data on each of these 97 stars in the HARPS sample from Tables 3 and 4 of their paper. We use the number of epochs per star, $N$, the overall uncertainties ($\sigma_{\rm tot}$) for each, including both ``internal'' ($\sigma_i$) and ``external'' ($\sigma_e$) errors, $\sigma_{\rm tot} \equiv \sqrt{\sigma_i^2+\sigma_e^2}$, and the mass of each star, $M_{\star}$ (see BX13 for a discussion of their uncertainties). We also obtain the time baseline for observations, $T$, for each star from the plots in their Figure 18. Since they do not report uncertainties in the host star mass estimates, we turn to the original reference for the method they use to compute the masses. \citet{2000A&A...364..217D} required that the stars they used to calibrate their mass-luminosity relationships have a mass accuracy of $\lesssim 10\%$, so we adopt uncertainties in the mass of the stars in the HARPS sample to be $10\%$. We use these data and the detection limits in figure 18 of BX13 to estimate their sensitivities and compute the expected number of planet detections and long-term RV trends as described in \S~\ref{subsec:method_application}.

We find the total expected number of planet detections by BX13 to be 
$N_{\rm det} = 1.4\pm 0.8$ and a lower limit on the number of trends they should see to be $N_{\rm t} = 2.1^{+1.2}_{-1.4}$, where the errors on these quantities are due to the uncertainties in the slope and normalization of our planetary mass function (see \S~\ref{subsec:method_application}). Our estimate of the number of trends is a lower limit because we are considering only populations of planets, whereas the RV survey could also be seeing trends due to distant stellar or brown dwarf companions. We bin the number of expected detections, trends, and total planets in decades of $m_p\sin{i}$ and $P$, similar to Table 11 in BX13, which we report in table \ref{tab:predicted_dets_trends}. For comparison, we also include the values reported by BX13.

In \citet{clanton_gaudi14a}, we determined that a fiducial RV survey (with $N=30$, $\sigma=4~{\rm m~s^{-1}}$, $T=10~$yr) should on average detect $0.049^{+0.46}_{-0.26}$ planets per star at a SNR of 5 or higher. If the sensitivities of BX13 for each star were equal to those of the fiducial survey, and if their sample covered the mass interval $0.07\leq M_{\star}/M_{\odot}\leq1.0$ in a log-uniform fashion (as was the case for our fiducial RV survey), we would have predicted that BX13 should have detected $4.9^{+4.6}_{-2.6}$ planets since their sample size is nearly $N_{\star}\sim 100~$ stars. This number is a factor $\sim 3.5$ larger than our final, detailed estimate. The difference arises from the fact that $\mathcal{Q}_{\rm min}=5$ for our fiducial survey, whereas the median value for HARPS is $\mathcal{Q}_{\rm min}\sim 10$, meaning our fiducial survey is overall more sensitive than HARPS. Our order of magnitude estimates turn out to be good enough to yield the right answer to within a factor of a few, but highlights the importance of understanding the detailed detection sensitivities of an entire sample to obtain accurate statistics.

\begin{table*}
\caption{\label{tab:predicted_dets_trends} Predicted detections and trends for the HARPS M dwarf survey (BX13), binned in $m_{\rm p}\sin{i}-P$ space. In each bin, $N_d$ is the number of predicted detections, $N_t$ is the number of predicted trends and $f$ is the derived planet frequency. The bold numbers are our results, while the unbolded values are those reported by BX13. There are no trend values for BX13 because it is not clear in which bins their reported trends lie (with the exception of Gl 832b, which we have included as a trend rather than a detection; see text). Uncertainties in our results are due to uncertainties in both the normalization and slope of the planetary mass function we adopt from the measurements by GA10 and \citet{2010ApJ...710.1641S}, respectively.}
\begin{tabular}{l|rrrrr}
\hline \hline
\multicolumn{1}{c|}{$m_{\rm p}\sin{i}$}  &  \multicolumn{5}{c}{Orbital Period [day]} \\
\multicolumn{1}{c|}{[M$_{\oplus}$]} &  1$-$10  & $10-10^2$  & $10^2-10^3$  & $10^3-10^4$ & $10^4-10^5$\\
        \hline
 & $N_d = \mathbf{0.0}, 0$& $N_d = \mathbf{0.0}, 0$& $N_d = \mathbf{(9.3^{+9.4}_{-9.28})E-3}, 0$& $N_d = \mathbf{0.013^{+0.011}_{-0.0126}}, 0$& $N_d = \mathbf{0.0d}, -$\\
$10^3-10^4$& $N_t = \mathbf{0.0}, -$& $N_t = \mathbf{0.0}, -$& $N_t = \mathbf{(4.5^{+5.3}_{-4.47})E-4}, -$& $N_t = \mathbf{0.093^{+0.066}_{-0.080}}, -$& $N_t = \mathbf{0.016^{+0.014}_{-0.0155}}, -$\\
 & $f = \mathbf{-}, <0.01$& $f = \mathbf{-}, <0.01$& $f = \mathbf{(1.0^{+1.1}_{-0.98})E-4}, <0.01$& $f = \mathbf{(1.2^{+8.4}_{-1.0})E-3}, <0.01$& $f = \mathbf{(3.6^{+3.4}_{-3.5})E-4}, -$\\
        \hline
 & $N_d = \mathbf{0.0}, 0$& $N_d = \mathbf{0.0}, 2$& $N_d = \mathbf{0.32^{+0.21}_{-0.24}}, 0$& $N_d = \mathbf{0.41^{+0.28}_{-0.29}}, 1$& $N_d = \mathbf{0.0}, -$\\
$10^2-10^3$& $N_t = \mathbf{0.0}, -$& $N_t = \mathbf{0.0}, -$& $N_t = \mathbf{0.012^{+0.011}_{-0.010}}, -$& $N_t = \mathbf{1.5^{+0.89}_{-1.0}}, 1$& $N_t = \mathbf{0.060^{+0.040}_{-0.046}}, -$\\
 & $f = \mathbf{-}, <0.01$& $f = \mathbf{-}, 0.02^{+0.03}_{-0.01}$& $f = \mathbf{(4.7^{+3.1}_{-3.4})E-3}, <0.01$& $f = \mathbf{0.038^{+0.023}_{-0.026}}, 0.019^{+0.043}_{-0.015}$& $f = \mathbf{(7.9^{+4.8}_{-5.4})E-3}, -$\\
        \hline
 & $N_d = \mathbf{0.0}, 2$& $N_d = \mathbf{(1.7^{+1.7}_{-1.6})E-4}, 0$& $N_d = \mathbf{0.28^{+0.14}_{-0.16}}, 0$& $N_d = \mathbf{0.31\pm 0.17}, 0$& $N_d = \mathbf{0.0}, -$\\
$10-10^2$& $N_t = \mathbf{0.0}, -$& $N_t = \mathbf{0.0}, -$& $N_t = \mathbf{0.023^{+0.013}_{-0.014}}, -$& $N_t = \mathbf{0.45\pm 0.24}, -$& $N_t = \mathbf{(1.0^{+1.3}_{-0.99})E-3}, -$\\
 & $f = \mathbf{-}, 0.03^{+0.04}_{-0.01}$& $f = \mathbf{-}, <0.02$& $f = \mathbf{0.020\pm 0.009}, <0.04$& $f = \mathbf{0.16^{+0.068}_{-0.072}}, <0.12$& $f = \mathbf{0.032^{+0.012}_{-0.014}}, -$\\
        \hline
 & $N_d = \mathbf{0.0}, 5$& $N_d = \mathbf{(2.9^{+2.9}_{-2.8})E-5}, 3$& $N_d = \mathbf{0.010\pm 0.007}, 0$& $N_d = \mathbf{(2.6^{+2.0}_{-2.3})E-3}, 0$& $N_d = \mathbf{0.0}, -$\\
$1-10$& $N_t = \mathbf{0.0}, -$& $N_t = \mathbf{0.0}, -$& $N_t = \mathbf{(3.7^{+1.9}_{-3.67})E-4}, -$& $N_t = \mathbf{(1.4^{+1.2}_{-1.39})E-3}, -$& $N_t = \mathbf{0.0}, -$\\
 & $f = \mathbf{-}, 0.36^{+0.24}_{-0.10}$& $f = \mathbf{-}, 0.52^{+0.50}_{-0.16}$& $f = \mathbf{0.080\pm 0.031}, -$& $f = \mathbf{0.64^{+0.25}_{-0.26}}, -$& $f = \mathbf{0.12^{+0.051}_{-0.049}}, -$\\
\hline\hline
\end{tabular}
\end{table*}

{\bf Detections: } Before we directly compare our predicted detections with the values reported by BX13, we first examine their reported detections. In the bin corresponding to $10^2 \leq m\sin{i}/ M_{\oplus} \leq 10^3$ and $10^3 \leq P/{\rm days} \leq 10^4$, they report the detection of two planets, Gl 832b and Gl 849b. They describe their data on these two planets in their \S~5.1. In the case of Gl 832b, they report that the HARPS data indicate a long-period RV variation at high confidence level, but with their data alone, they cannot uniquely determine the Keplerian orbit and thus are unable to confirm the planetary nature of Gl 832b. Only when they combine the HARPS data with the AAT data, are they able to refine the orbit of the planet and confirm its planetary nature. Thus, we argue that the HARPS survey sample should not include the detection of Gl 832b when determining planet frequencies from their survey. In the case of Gl 849b, the HARPS data confirms it as a Jupiter-mass companion. When they combine Keck RVs for this planet, they report that a single planet is not enough to explain the RV variation, but since they are able to identify the companion as a planet with HARPS data alone, this planet should be included in the sample. Therefore, the number of detections in this bin of Table 11 of BX13 should to be one, rather than two, and the planet frequency here should be $f=0.019^{+0.043}_{-0.015}$. However, since the HARPS data alone confirm long-term variation, we include Gl 832b as an identified trend by their survey.

In particular, we focus on comparing our predictions for planet detections and trends with the actual numbers reported by BX13 for orbital periods longer than $\sim 100~$days. Microlensing surveys have little or no sensitivity to shorter orbital periods and thus we are unable to compare with in these regions where there is no overlap between microlensing and RV. We predict that BX13 should detect a total of $N_{\rm det} = 1.4\pm 0.8$ planets. The majority of these predicted planet detections for HARPS lie in four bins (see table~\ref{tab:predicted_dets_trends}). The largest amount of predicted planet detections, with $N_d=0.41^{+0.38}_{-0.18}$, lie in the $10^2\leq m_p\sin{i}/M_{\rm Jup}\leq 10^3$ and $10^3\leq P/{\rm days}\leq 10^4$ bin. The only reported planet detection by BX13 falls into this bin (Gl 849b) with a mass of $m_p\sin{i}=372\pm 19~M_{\oplus}$ and an orbital period of $P=2165\pm 132~$days. The other three bins within which we predict a significant amount of planet detections include $10^2\leq m_p\sin{i}/M_{\rm Jup}\leq 10^3$ and $10^2\leq P/{\rm days}\leq 10^3$ with $N_d=0.32^{+0.29}_{-0.16}$, $10\leq m_p\sin{i}/M_{\rm Jup}\leq 10^2$ and $10^3\leq P/{\rm days}\leq 10^4$ with $N_d=0.31^{+0.21}_{-0.13}$, and finally $10\leq m_p\sin{i}/M_{\rm Jup}\leq 10^2$ and $10^2\leq P/{\rm days}\leq 10^3$ with $N_d=0.28^{+0.18}_{-0.12}$. The fact that BX13 do not report any planet detections in these three bins is consistent with our predictions since the Poisson probabilities of detecting zero planets, assuming the predicted number of detections is equal to the mean number of planets residing in these bins such that $P(0)=e^{-N_d}$, are $0.74\pm 0.12$, $0.75^{+0.16}_{-0.17}$, and $0.76^{+0.12}_{-0.11}$, respectively.

In summary, we predict that the HARPS survey should find about one planet with a period right at the edge of the survey duration and indeed BX13 report the detection of such a planet (Gl 849b). Thus, consistency between microlensing and radial velocity surveys in the region of planet parameter space in which they overlap implies that the giant planet frequencies inferred from the two types of surveys are in fact consistent. We conclude that RV surveys are detecting only the high-mass end of the population of giant planets inferred by microlensing, leading to their underestimate of the total giant planet frequency around M dwarfs.

{\bf Trends: } In our approximation, we expect planets to be identified as long-term RV drifts when they have periods greater than the time baseline of observations of their host star, i.e. $P>T$, and produce detectable signals, i.e. lying on or above the detection limit curve for their host star (as exemplified in figure \ref{fig:marg_dist}). In the limit $P\gg T$, the RV trends will be basic, linear accelerations, the slope of which depends on the phase covered by the actual observations. However, when $P$ is just larger than $T$, by our approximation such a planet will also be considered as a trend, but will exhibit more complex variation than a linear trend. We compute the RV accelerations for our predicted trend-producing planets by multiplying the maximum possible slope, $2\pi K/P$, by a factor $\cos{\phi}$, where $\phi$ is the phase angle at the time of observation, randomly and uniformly drawn between $[0,2\pi)$. We ignore the eccentricity in computing the slopes and make the approximation $P\gg T$.

Under these assumptions, we predict that the HARPS M dwarf survey should find at least one or two trends ($N_{\rm t} = 2.1^{+1.2}_{-1.4}$), with a median RV acceleration and 68\% confidence interval of $7.9_{-5.8}^{+19.}~{\rm m~s^{-1}~yr^{-1}}$, most likely in the bin with $10^3\leq P/{\rm days}\leq 10^4$ and $10^2\leq m_p\sin{i}/M_{\oplus}\leq 10^3$ (there is expected to be $1.5^{+1.3}_{-0.6}$ RV trends due to planets in this bin as shown in table~\ref{tab:predicted_dets_trends}). As discussed above, Gl 832b falls in this bin with a reported acceleration of $5.198~{\rm m~s^{-1}}$. Indeed, the RV time series for this star (shown in figure 3 of BX13) does exhibit more complex variability than a simple linear trend. BX13 report additional long-term RV trends in their sample. The largest, statistically significant RV acceleration (i.e. with a  false alarm probability (FAP) less than 0.01) reported by BX13 is $-9.616~{\rm m~s^{-1}~yr^{-1}}$ from the star Gl 849 (MB14 also detect RV acceleration of this star). They report a total of 15 stars to have RV slopes with FAP$<0.01$, with a median magnitude of $2.65~{\rm m~s^{-1}~yr^{-1}}$. Of these 15 stars, the report only five of them to have ``smooth'' RV drifts, namely LP 771-95A, Gl 367, Gl 618A, Gl 680, and Gl 880, while the rest exhibit more complex variability. The median magnitude of these ``smooth'' RV accelerations is $3.20~{\rm m~s^{-1}~yr^{-1}}$.

\begin{figure}[h!]
\epsscale{1.1}
\plotone{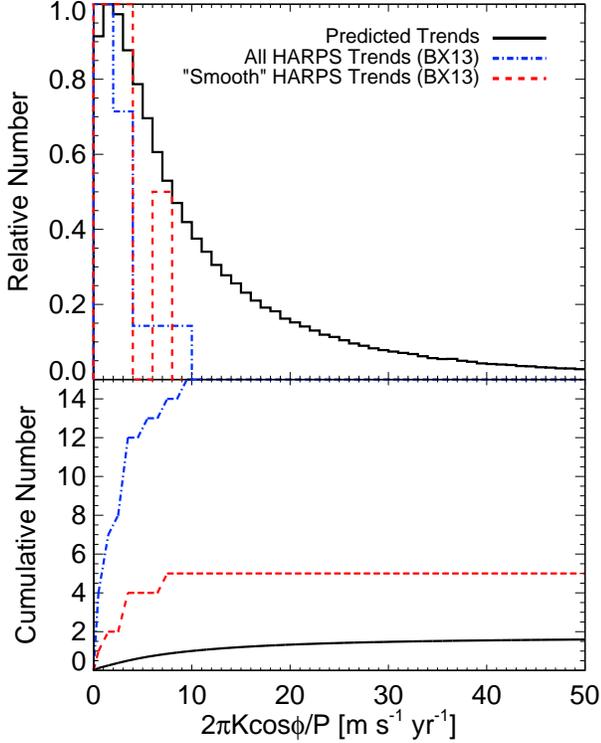}
\caption{The top panel shows the relative number of long-term RV trends for the actual HARPS sample and our predicted sample. The blue dot-dashed lines include the stars BX13 report to have significant RV trends (with FAP$<0.01$) and the red dashed lines are a subset of these stars for which BX13 report smooth RV variation. The black lines are our predicted trends which are computed as $2\pi K\cos{\phi}/P$ for the systems we expect to show up with trends. The bottom panel shows the cumulative distribution functions of these distributions. We perform K-S tests and find that our predicted distribution of trends is inconsistent with the distribution of all RV trends (yet not necessarily the subset of smooth trends), suggesting that a majority of the trends identified in the HARPS sample, if arising from companions, are due to more distant and more massive stellar or brown dwarf companions, or planets to which microlensing is not sensitive.
  \label{fig:trends}}
\end{figure}

Figure \ref{fig:trends} shows the histograms and CDFs of all trends and the smooth trends reported by BX13, along with the distribution of drifts that we predict. We perform a two-sample Kolmogorov-Smirnov (K-S) test between our predicted distribution of RV trends and that of all 15 significant trends from HARPS and find a $D$-statistic of 0.52 with probability $P\left(D\right)=2.8\times10^{-3}$, demonstrating that the two distributions are inconsistent. We also perform a two-sample K-S test between our predicted distribution and that of just the 5 significant, smooth trends found in the HARPS sample, which yields $D=0.52$ with probability $P\left(D\right)=0.084$.

We can explain the RV accelerations BX13 detect from Gl 832b and Gl 849c as arising from planetary companions predicted by microlensing. In the next section, we discuss how MB14 are able to constrain the mass of Gl 849c to be $m_p\sin{i}=0.70\pm 0.31~M_{\rm Jup}$ and its orbital period to be $19.3^{+17.1}_{-5.9}$~years by measuring the rate of change in RV acceleration, or the ``jerk.'' This most likely places Gl 849c into the same bin of mass and period as Gl 832b, where we predict $1.5^{+1.3}_{-0.6}$. However, the remaining 13 RV drifts are inconsistent with the hypothesis that they are caused by planetary companions analogous to the population inferred from microlensing. MB14 suggest that at least two of the trends detected by BX13, those of Gl 250B and Gl 618B, can be attributed to long-period binary companions. It is unclear if the remainder of the RV trends are due to planets beyond the sensitivity of current microlensing surveys, stellar or brown dwarf binary companions, or even magnetic activity \citep[e.g.][]{1988lsla.book.....G,2012A&A...541A...9G}.

\begin{figure}[h!]
\epsscale{1.1}
\plotone{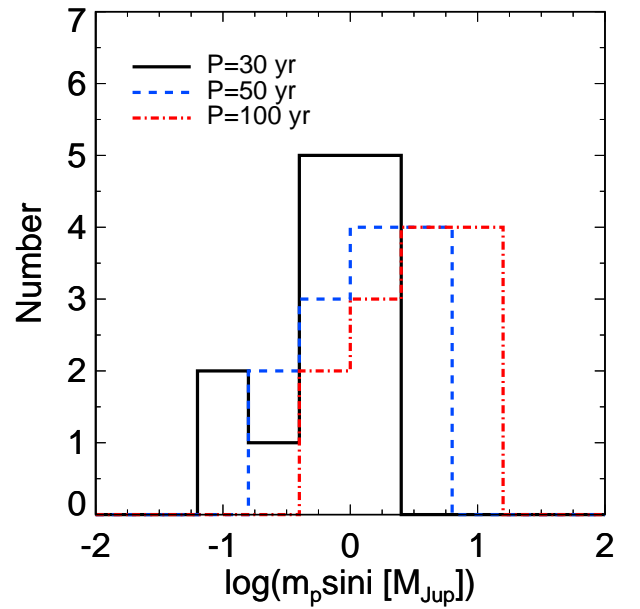}
\caption{The minimum companion masses required to produce the 13 unexplained long-term RV trends observed by BX13 in the HARPS M dwarf sample assuming the source of these trends is at a given period (see equation~\ref{eqn:min_trend_mass}). The black histogram represents the minimum required masses if the companions have orbital periods of 30~years. The blue, dashed histogram and the red, dot-dashed histogram represents these companions at orbital periods of 50 and 100~years, respectively.
  \label{fig:possible_trends_masses}}
\end{figure}

We can assess the plausibility that the measured trends are due to planetary mass companions that are at periods outside those for which microlensing is sensitive. If we let $a_{\rm t}$ be the magnitude of a given trend measured by BX13, then setting $2\pi K\cos{\phi}/P=a_{\rm t}$, substituting for $K$ using the standard velocity semi-amplitude equation, and solving for $m_p\sin{i}$ yields the minimum companion mass required to produce the observed trend as a function of orbital period
\begin{align}
    m_p\sin{i} = & {} \left(\frac{P}{2\pi}\right)^{4/3}G^{-1/3}M_{\star}^{2/3}\frac{a_{\rm t}}{\cos{\phi}} \\
    = & {}\; 0.44~M_{\rm Jup} \left(\frac{a_{\rm t}}{1~{\rm m~s^{-1}~yr^{-1}}}\right)\left(\frac{P}{30~{\rm yr}}\right)^{4/3} \nonumber \\
    & {} \times\left(\frac{M_{\star}}{0.5~M_{\odot}}\right)^{2/3}\left(\frac{1}{\cos{\phi}}\right) \; . \label{eqn:min_trend_mass}
\end{align}
Using equation~\ref{eqn:min_trend_mass}, we plot the minimum required companion mass to yield the measured RV accelerations reported by BX13 for the 13 unexplained trends in figure~\ref{fig:possible_trends_masses}. We assume that $\cos{\phi}=1$ in our calculations because the exact orbital phase during observations is unknown; any other value of $\cos{\phi}$ would serve to increase the required companion mass, so this assumption assures we are indeed estimating the minimum required companion mass. We plot these values assuming the companions are at orbital periods of 30, 50, and 100~years. As we have previously shown, a planet with an orbital period of about $P\sim 30~$years, which corresponds to a projected separation of roughly 2.5 times the Einstein radius of the typical lens, is just beyond the sensitivity of microlensing surveys. The minimum required companion masses at all periods are consistent with giant planets ($m_p\sin{i}>0.1~M_{\rm Jup}$), with just one exception. BX13 report the measurement of a $0.206~{\rm m~s^{-1}~yr^{-1}}$ RV acceleration of Gl 431.1, which has a minimum required companion mass of roughly $27~M_{\oplus}$ if it orbits at a period of 30~years. Thus, if giant planets are common at orbital periods beyond $\sim 30~$years, it is plausible that these are the source of the majority of the long-term RV trends measured by BX13 in the HARPS M dwarf sample. However, we note that there are significantly less trends reported by MB14 for the CPS M dwarfs despite having a larger sample size than HARPS.

\subsection{Comparison with CPS Planet Detections}
\label{subsec:johnson_comparison}
MB14 provide basic parameters for each of the 111 M dwarfs in their sample (which we describe in \S~\ref{subsec:johnson_sample}), including the stellar mass, number of RV measurements, time baseline of observations, and average RV precision. Since the CPS team has not yet determined individual RV detection sensitivities for each of their stars, we make a very rough estimate of their sensitivity by matching CPS stars with those from the HARPS sample with similar systematics as described in \S~\ref{subsec:method_application}. We then determine the expected number of planet detections and long-term RV trends the CPS team should see in the same manner as we did for the HARPS sample.

{\bf Detections: }We predict a total of $N_{\rm det}= 4.7^{+2.5}_{-2.8}$ detected planets with periods longer than $10^2~$days from the CPS M dwarf sample, and indeed this sample has yielded 4 such planets. We expect $2.2^{+1.4}_{-1.5}$ of our predicted planet detections to have a mass between $10^2-10^3~$M$_{\oplus}$ and a period between $10^3-10^4~$days. Two of the CPS detections, Gl 179b \citep{2010ApJ...721.1467H} and Gl 849 \citep{2006PASP..118.1685B}, lie in this bin. The other two CPS detections, Gl 317b \citep{2007ApJ...670..833J} and Gl 649b \citep{2010PASP..122..149J}, lie in the mass range $10^2\leq m_p\sin{i}/M_{\oplus}\leq 10^3$ and the period range $10^2\leq P/{\rm days}\leq 10^3$. Of our predicted planets, we expect $0.46^{+0.30}_{-0.34}$ to lie in this bin. If this number is indeed true number of planets in this bin, then the Poisson probability of detecting two planets is $0.19^{+0.07}_{-0.08}$, which we consider to be marginally significant. However, as we discuss in \S~\ref{sec:synthesis}, the sensitivity of microlensing falls off towards shorter periods in this bin, while the sensitivity of RV surveys decreases towards longer periods. We therefore expect the planet frequency in this bin to be larger than the value we predict from microlensing in this paper, so it is not surprising that we under-predict the number of planet detections in this period range. Of the remaining predicted planet detections, we expect $1.4^{+0.72}_{-0.74}$ planet detections with $10\leq m_p\sin{i}/M_{\oplus}\leq 10^2$ and $10^3\leq P/{\rm days}\leq 10^4$, $0.44^{+0.21}_{-0.23}$ detections with $10\leq m_p\sin{i}/M_{\oplus}\leq 10^2$ and $10^2\leq P/{\rm days}\leq 10^3$, and $0.12^{+0.09}_{-0.10}$ detections with $10^3\leq m_p\sin{i}/M_{\oplus}\leq 10^4$ and $10^3\leq P/{\rm days}\leq 10^4$. There are no CPS detections in these bins, the Poisson probabilities for which are $0.31\pm 0.20$, $0.67^{+0.15}_{-0.14}$, and $0.90^{+0.11}_{-0.09}$, respectively, assuming that the true number of planets in these bins are the predicted values.

{\bf Trends: }We predict that the CPS M dwarf sample should see a total of $N_{\rm t}= 1.8^{+1.1}_{-1.2}$ long-term RV drifts due to giant planets on long-period orbits. Of these, we predict $1.1^{+0.69}_{-0.75}$ will be due to a giant planet with $0.31\lesssim m_p\sin{i}/M_{\rm Jup}\lesssim 3.1$ and $2.7\lesssim P/{\rm yr}\lesssim 27$. There are four other bins that we predict to harbor a significant source of RV trends in the CPS sample: between $0.31\leq m_p\sin{i}/M_{\rm Jup}\leq 3.1$ and $27\leq P/{\rm yr}\leq 270$ we predict $0.30^{+0.19}_{-0.22}$ trends, between $0.031\leq m_p\sin{i}/M_{\rm Jup}\leq 0.31$ and $2.7\leq P/{\rm yr}\leq 27$ we predict $0.28\pm 0.15$ trends, between $3.1\leq m_p\sin{i}/M_{\rm Jup}\leq 31$ and $2.7\leq P/{\rm yr}\leq 27$ we predict $0.081^{+0.065}_{-0.073}$ trends, and between $3.1\leq m_p\sin{i}/M_{\rm Jup}\leq 31$ and $27\leq P/{\rm yr}\leq 270$ we predict $0.042^{+0.038}_{-0.040}$ trends.

MB14 report a total of four measured RV accelerations. Of these, that of Gl 849 exhibits significant curvature (or ``jerk''), allowing for constraints on the mass and period of the long-period companion. They find a median minimum mass of $m_p\sin{i}=0.70\pm 0.31~M_{\rm Jup}$ and a median period of $19.3^{+17.1}_{-5.9}~$years. Although MB14 are able to place constraints on the companion properties for this measured trend (and imaging rules out stellar mass companions), in our simulation this planet would be counted as a trend. The mass and period most likely place it in the bin we predict the most trends to lie. The weak constraints on the orbital period could scatter this trend into the next higher period bin, which happens to be another bin for which we predict a significant number of trends.

The remaining three other stars for which MB14 measure significant RV accelerations are Gl 317 ($2.51\pm 0.62~{\rm m~s^{-1}~yr^{-1}}$), Gl 179 ($-1.17\pm 0.29~{\rm m~s^{-1}~yr^{-1}}$), and Hip 57050 ($1.39\pm 0.39~{\rm m~s^{-1}~yr^{-1}}$). Imaging with NIRC2 (instrument PI: Keith Matthews) using the AO system at the W. M. Keck Observatory \citep{2000PASP..112..315W} in the $K'$ or $K_s$ filters rule out most stellar-mass companions and some brown dwarfs as the source of these trends. The low inferred brown dwarf frequency around M dwarfs from \citet{2012AJ....144...64D} and similarly low frequency of brown dwarf companions to FGK stars from \citet{2009ApJS..181...62M} lead MB14 to conclude that these trends are probably due to giant planets. However, they do mention that their imaging of Hip 57050 is only complete at separations smaller than 1 arcsecond ($r_{\perp}\approx 11~$AU), leaving some parameter space for a low mass M dwarf companion to be the cause of the RV acceleration.

We predict a trend that is consistent with that caused by Gl 849, but overall our numbers seem to be marginally consistent with the four observed RV accelerations by MB14, if they are indeed due to planetary companions. The Poisson probability of detecting four trends when the true mean is $N_t=1.8^{+1.1}_{-1.2}$ is $0.036^{+0.11}_{-0.033}$. If MB14 have misclassified one of their detected trends, and turns out to be due to a brown dwarf companion rather than a planetary companion, then the Poisson probability of detecting 3 trends if the true mean is $N_t=1.8^{+1.1}_{-1.2}$ increases to $0.10^{+0.11}_{-0.08}$. 

As we did for the BX13 trends, we can compute the minimum companion mass required to produce the trends MB14 measure for Gl 317, Gl 179, and Hip 57050 using equation~\ref{eqn:min_trend_mass}. At a period of 30~years, the minimum required companion mass for these stars is $1.0~M_{\rm Jup}$, $0.47~M_{\rm Jup}$, and $0.55~M_{\rm Jup}$, respectively. At a period of 50~years, we calculate $2.0~M_{\rm Jup}$, $0.92~M_{\rm Jup}$, and $1.1~M_{\rm Jup}$, respectively. At 100~years, we find $5.0~M_{\rm Jup}$, $2.3~M_{\rm Jup}$, and $2.8~M_{\rm Jup}$, respectively. The companions responsible for producing these long-term trends MB14 measure could be giant planet planets and would be beyond the sensitivity of current microlensing surveys. We note that although our inferred frequency is consistent with that of MB14, it is nevertheless a median factor of 2.2 ($0.22-8.8$ at 95\% confidence) times smaller, potentially due to the fact that microlensing is missing such a population of very long-period super-Jupiters, which is being inferred by MB14 by these trends that microlensing does not predict.  In fact, if MB14 were to ignore these three trends, we expect they would infer a frequency nearly identical to ours.

\begin{figure}[h!]
\epsscale{1.1}
\plotone{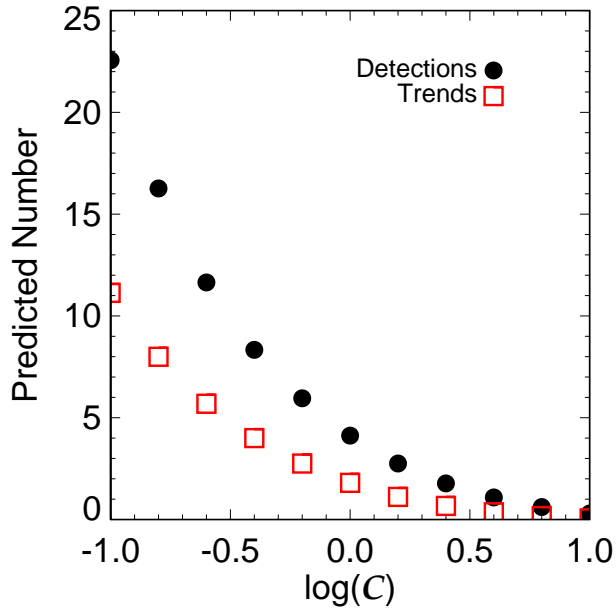}
\caption{The numbers of predicted detections and trends as a function of the SNR scale factor, $\mathcal{C}$. Since we do not have detailed detection sensitivities for each star in the CPS sample, we estimate their detection limits by assuming that stars in the CPS sample with similar systematics to stars in the HARPS sample have the same sensitivities (see \S~\ref{subsec:method_application}). This plot shows that our predicted number of detections and trends depends sensitively on the detection limits we assume due to the steeply declining mass function inferred from microlensing surveys.
  \label{fig:nd_nt_scaling}}
\end{figure}

One caveat with this comparison is that we do not have the actual detection limits for each star in the CPS sample so we are forced to estimate them by matching to stars in the HARPS M dwarf sample with similar systematics. Due to the steep planetary mass function inferred from microlensing, the numbers of predicted detections and trends are very sensitive to the detection limits. The right panel in figure \ref{fig:marg_dist} showing the distribution of $K$ marginalized over all $P$ reflects this steep mass function. In order to make more robust predictions for the CPS M dwarf sample examined by MB14, we would need more accurate sensitivity estimates. In order to illustrate this point, we assume that the overall distribution of $\mathcal{Q}_{\rm min}$ remains the same, but multiply each by a constant SNR scale factor, $\mathcal{C}$, to determine the detection limits for the CPS M dwarf sample. We then calculate the total predicted numbers of planet detections and trends. We plot $N_d$ and $N_t$ as a function of $\mathcal{C}$ in figure~\ref{fig:nd_nt_scaling}. Increasing $\mathcal{Q}_{\rm min}$ by a factor of 2 results in roughly 1.8 fewer detections and 0.9 fewer trends, while decreasing $\mathcal{Q}_{\rm min}$ by a factor of 2 results in roughly 3 more detections and 1.6 more trends.

\subsection{Additional Simulations and Results}
\label{sec:additional_simulations}
We ran additional simulations where we altered the systematics of the HARPS and CPS surveys in order to see how the numbers of predicted detections and trends would change if the time baselines for each star were increased, or if they were able to reduce both internal and external noise sources. We ran three different tests where we 1) doubled the time baseline, $T$, of observations for each star, 2) fixed measurement errors at $\sigma = \left(\sigma_i^2+\sigma_e^2\right)^{1/2}=1~{\rm m~s^{-1}}$, and 3) both doubled $T$ and fixed $\sigma = 1~{\rm m~s^{-1}}$. The results from each of these simulations for the HARPS survey are as follows: 1) $N_{\rm det} = 2.0\pm 1.4$, $N_{\rm t} = 1.2\pm 0.9$, 2) $N_{\rm det} = 3.7\pm 2.7$, $N_{\rm t} = 6.3\pm 4.5$, and 3) $N_{\rm det} = 6.7\pm 4.8$, $N_{\rm t} = 4.5\pm 3.2$. As expected, we find that increasing the duration of observations and reducing the uncertainties increases the numbers of predicted detections and trends. For the CPS sample, we find the results: 1) $N_{\rm det} = 4.9\pm 3.5$, $N_{\rm t} = 0.60\pm 0.43$, 2) $N_{\rm det} = 11.\pm 8$, $N_{\rm t} = 4.6\pm 3.3$, and 3) $N_{\rm det} = 14.\pm 10$, $N_{\rm t} = 2.1\pm 1.5$. Doubling observation times does not double the expected number of detections, as the median time baseline for the CPS sample is over 10 years, whereas the median period for planets found by microlensing surveys is about $9.4~$years (see Figure \ref{fig:marg_dist}).

At least in terms of orbital period, the CPS survey is sensitive to a majority of the population of planets inferred from microlensing surveys. In the case of both RV surveys, decreasing measurement uncertainties greatly increases the number of expected RV planet detections at a range of orbital periods, but peaking near the edge of their survey durations. Thus, if RV surveys hope to detect the entire population of giant planets inferred by microlensing, rather than just the high-mass end, the typical measurement uncertainties need to be reduced by a factor of a few to cut further into the steep planetary mass function.

\subsection{Properties of Planets Accessible to Microlensing and RV}
\label{subsec:planet_props}
In addition to computing the number of planet analogs to which RV surveys are sensitive, we are also interested in the properties of such planets. In the appendix, we examine the distributions of microlensing and orbital parameters for the planets we predict will show up as detections and long-term RV trends in the HARPS sample to determine if there is a subset of the planet population inferred from microlensing towards which RV surveys are particularly sensitive.

Not surprisingly, we find that the planets we predict HARPS will detect is sensitive to the distribution of projected separations, $s$, preferring small values of $s$, and preferring higher values of the planet to host star mass ratio, $q$. We also find that predicted detections have a slight bias against lens distances at and near the halfway point between the Earth and the source, where the Einstein radius is maximized. This is a reflection of the fact that the RV signal decreases with increasing orbital separation, as well as the fact that the median time baseline for stars in the HARPS sample is shorter than the median period of the entire population of microlensing planets by a factor of $\sim 2$. Additionally, there is a preference for planet detections around more massive hosts, even at fixed $q$. However, we find that there is no significant preference for RV planet detections of analogs to the planets found around bulge or disk lenses by microlensing (assuming planets are equally common around all stars regardless of their location in the Galaxy). See the appendix for additional discussion.

\section{Uncertainties}
\label{sec:uncertainties}
\subsection{Normalization and Slope of the Microlensing Mass-Ratio Function}
\label{subsec:norm_slope_unc}
The main sources of uncertainty in our calculations are the uncertainties in the microlensing measurements of the normalization and slope of the planetary mass function \citep[GA10;][]{2010ApJ...710.1641S}. The quoted uncertainties of our results throughout this paper are due to these sources. See \S~\ref{subsec:method_application} and \citet{clanton_gaudi14a} for a description of how these uncertainties are propagated.

There is another source of uncertainty, which we mention here but do not explicitly include in our final results.  This stems from our assumption that the distribution function of companions $d^2N_{\rm pl}/(d\log{q}~d\log{s})$ is invariant in mass ratio, $q$, rather than planet mass, $m_p$.  Microlensing surveys are currently unable to distringuish between these two assumptions. Therefore, the planet frequency we infer for planets of a given $m_p$ depends on the primary mass.  We have adopted a typical primary mass for the microlensing sample of $M_l\sim 0.5~M_{\odot}$.  On the other hand, the median stellar mass of the HARPS M dwarf sample (BX13) is $0.3~M_{\odot}$ and that of the CPS M dwarf sample (MB14) is $0.41~M_{\odot}$. Therefore, our assumption of a fixed distribution function in $q$ means that we are assigning a lower planet frequency at fixed planet mass for the BX13 and MB14 samples than for the microlensing sample.  However, the mass distribution and typical mass of the microlensing sample is uncertain, and values as low as $0.3~M_{\odot}$ are possible.  Had we adopted lower values, our inferred frequencies for the HARPS and CPS sample would be higher. To estimate the level of this effect, we integrate our planetary mass-ratio function over the mass interval $1\leq m_p/M_{\rm Jup}\leq 13$ assuming a host mass of $M_{\star} = 0.5~M_{\odot}$ and divide by the mean value of this same integral calculated for the host star masses of each of the stars in the  HARPS sample.  We then repeat this exercise for the CPS sample. We find frequencies that are factors of 1.4 and 1.2 times higher, respectively, indicating that the actual frequencies we infer could be up to $\sim 40\%$ higher for the HARPS sample and up to $\sim 20\%$ for the CPS sample. Given the fact that the uncertainties on our final results due to the slope and normalization of our planetary mass-ratio function are typically around the $\sim 50\%$ level, there are some cases where this effect could be significant.

\subsection{Galactic Model and Microlensing Parameter Distribution}
\label{subsec:gal_model_mlens_param_unc}
There is also some degree of unquantified uncertainties due to our choice of priors on the planet and host star properties of our simulated sample (e.g. priors on lens masses and distances, planetary orbital parameters). However, we expect any such errors to be subdominant because we are mostly able to reproduce the observed distributions of host star parameters from the actual microlensing sample by appropriately weighting by the event rate, with the single possible exception of the distribution of lens distances. We describe in great detail the comparison of the distribution of such parameters between our simulated sample and the actual microlensing sample of GA10 in \citet{clanton_gaudi14a}.

\subsection{Contamination from FGK Stars and Remnants}
\label{subsec:contam_unc}
There could be unquantified sources of error in our analysis related to differences between the microlensing and RV samples. For example, microlensing is only able to measure lens masses for a subset of all events. While each of the planet-hosting lenses in the GA10 sample have mass measurements (or at least mass upper limits), it could be the case that a fraction of the lenses included in the GA10 sample are not actually M dwarfs, but are instead stellar remnants (white dwarfs, neutron stars, or black holes) or even K and G stars. \citet{2000ApJ...535..928G} estimated that $\sim 20\%$ of detected microlensing events are due to remnants that are completely unrecognizable from their timescale distribution. Consequently, we expect the resultant uncertainty to be small in comparison to the Poisson error on the number of planet detections included in the GA10 study and thus not a significant source of error in our analysis.

\subsection{Differences in the Metallicity Distribution of RV and Microlensing Hosts}
\label{subsec:metallicity_unc}
In general, any Galactic gradient of properties that affect planet frequency could affect our results. The most obvious of such properties is the Galactic metallicity gradient \citep[see e.g.][]{2012ApJ...746..149C,2013arXiv1311.4569H}. While RV surveys of M dwarfs are limited to targets within tens of parsecs, microlensing probes stellar hosts much further away and towards the Galactic center, at distances of a few to several kiloparsecs. Microlensing also probes stars in the Galactic bulge, which may not form giant planets \citep[e.g.][]{2013MNRAS.431...63T}. The metallicities of the disk stars in microlensing samples are therefore expected to be enhanced relative to those monitored by RV. RV surveys have shown a strong correlation between metallicity and planet frequency over a wide range of metallicities \citep[e.g.][JJ10, MB14]{2005ApJ...622.1102F}, and thus the Galactic metallicity gradient has been hypothesized to be the cause of the difference in inferred giant planet frequency around M dwarfs between microlensing and RV surveys. JJ10 found the empirical relation between giant planet occurrence, stellar mass, and metallicity
\begin{align}
    f\left(M_{\star}, {\rm [Fe/H]}\right) = & {} \left(0.07\pm 0.01\right)\left(M_{\star}/M_{\odot}\right)^{1.0\pm 0.3} \nonumber \\
    & {} \times 10^{(1.2\pm 0.2){\rm [Fe/H]}}\;\label{eqn:jj_metallicity_corr}
\end{align}
for giant planets ($K>20~{\rm m~s^{-1}}$) on orbits within $a<2.5~$AU by analyzing the full CPS sample, which includes 1194 stars in the mass interval $0.2<M_{\star}/M_{\odot}\lesssim 2.0$ and the metallicity interval $-1.0<{\rm [Fe/H]}<+0.55$. Examining just the CPS M dwarfs, MB14 find the relation
\begin{align}
    f\left(M_{\star}, {\rm [Fe/H]}\right) = & {} 0.039^{+0.056}_{-0.028}\left(M_{\star}/M_{\odot}\right)^{0.8^{+1.1}_{-0.9}} \nonumber \\
    & {} \times 10^{(3.8\pm 1.2){\rm [Fe/H]}}\;,\label{eqn:mb_metallicity_corr}
\end{align}
for planets with masses $1<m_p\sin{i}/M_{\rm Jup}<13$ on orbits within $a<20~$AU, which has a significantly steeper scaling with metallicity than the JJ10 result. This implies that the dependence of the frequency of Jupiter and super-Jupiter mass planets on host metallicity is much steeper for M dwarfs than for higher mass stars. On the other hand, \citet{2013A&A...551A..36N} find a more shallow metallicity dependence for Jovian hosts of
\begin{equation}
    f({\rm [Fe/H]}) = (0.02\pm 0.02)\times 10^{(1.97\pm 1.25){\rm [Fe/H]}}\; \label{eqn:harps_metallicity_corr}
\end{equation}
by examining the HARPS M dwarf sample.\footnote{Although \citet{2013A&A...551A..36N} do not specify a period range over which this relation is valid, we can reasonably assume it holds for periods less than a couple thousand days, which is roughly the median time baseline of observations for the HARPS M dwarf sample (BX13).} These authors also analyze a combined HARPS and CPS M dwarf data set and report
\begin{equation}
    f({\rm [Fe/H]}) = (0.03\pm 0.02)\times 10^{(2.94\pm 1.03){\rm [Fe/H]}}\; \label{eqn:harps_cps_metallicity_corr}
\end{equation}
for Jovian hosts from the combined data set. MB14 acknowledge the shallower dependence on metallicity reported by the \citet{2013A&A...551A..36N} study and attribute it to their inclusion of a sub-Jupiter mass planet in their sample of Jovian hosts, which happens to orbit a star with a metallicity of ${\rm [Fe/H]=-0.19\pm 0.08}$. MB14 further emphasize the fact that there are no planets with $m_p\sin{i}>1~M_{\rm Jup}$ orbiting M dwarfs with measured metallicities below $+0.08~$dex in either the HARPS or CPS samples.

We would like to know what these relations between planet frequency and metallicity imply for the frequency of giant planets expected from the microlensing sample. We therefore apply these relations to a simulated microlensing host star sample with a stellar mass distribution similar to that expected for actual microlensing samples, covering the range $0.07\leq M_l/M_{\odot}\leq 1.0$ in a log-uniform fashion (see \citealt{clanton_gaudi14a} for details on creating such a sample). The remaining task is to determine the metallicities of the stars in our simulated sample.

Actual microlensing samples are mostly comprised of low-mass and distant (and thus faint, typically with $V\gtrsim 18$) stars, the light from which is often blended with that of the source and perhaps also nearby stars due to crowded fields and limited seeing from ground-based observations. Metallicity measurements are therefore out of reach with current technology and the metallicity distribution of the microlensing sample remains unknown. Instead, we estimate the metallicity distribution of our simulated sample using the recent Galactic metallicity maps from the SDSS-III APOGEE experiment \citep{2013arXiv1311.4569H} for our disk lenses, and the bulge metallicity distribution function (MDF) derived from a sample of microlensed dwarfs and subgiants \citep{2013A&A...549A.147B} for our bulge lenses. As we demonstrate in \citet{clanton_gaudi14a}, the parameter distributions (e.g. $M_l$, $t_E$) of our simulated sample basically match those of the GA10 sample (except for lens distances --- we will come back to this later in the section), and thus we expect the metallicity distribution for the actual microlensing sample to be roughly similar to that which we derive here.

We determine the metallicities of our simulated disk lenses as follows. Table 2 of \citet{2013arXiv1311.4569H} lists the parameters of their fits to the measured metallicities as a function of height above the plane, $z$, and Galactocentric radius, $R$. We model the median metallicities of disk stars as a function of $R$ (in several bins of $|z|$, as in \citealt{2013arXiv1311.4569H}) using these linear fits. At a given $R$, we assume the distribution in metallicity about these median values is a Gaussian with a standard deviation of $0.2~$dex, which is equal the measured spread these authors report. We then assign our disk lenses a random metallicity drawn from a Gaussian constructed in the above manner. The Galactocentric radius of a given disk lens, with distance $D_l$ from Earth and at Galactic longitude and latitude $(l,b)$, is $R=(x^2+y^2)^{1/2}$, where $x=R_0-D_l\cos{l}\cos{b}$ and $y=D_l\sin{l}\cos{b}$, and where we take $R_0=8~$kpc as the Solar radius. The height above the Galactic disk of a given lens is given by $z=D_l\sin{b}$. We set a maximum possible metallicity of ${\rm [M/H]}=0.6~$dex as there are no measurements of stellar metallicities larger than this value in \citet{2013arXiv1311.4569H}.

We assign each of our bulge lenses a random metallicity from the MDF shown in figure 12a of \citet{2013A&A...549A.147B}, regardless of the location of the event, $(l, b)$, since \citet{2013A&A...549A.147B} do not find statistically significant differences in the metallicity distributions of stars closer to ($\left|b\right|\leq 3^{\circ}$) or farther from ($\left|b\right|> 3^{\circ}$) the Galactic plane nor in the metallicity distributions of stars closer to ($\left|l\right|\leq 2^{\circ}$) or farther from ($\left|l\right|> 2^{\circ}$) the Galactic center.

\begin{figure}[h!]
\epsscale{1.1}
\plotone{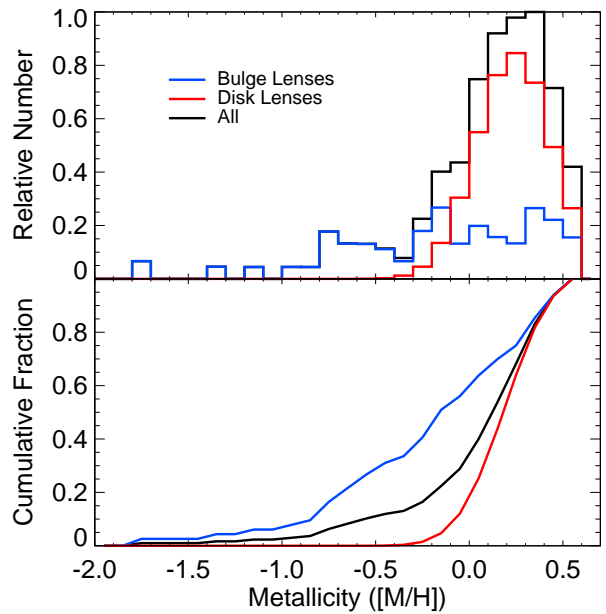}
\caption{The top panel shows the relative number of lens star metallicities for our simulated microlensing sample. The blue and red lines represent the metallicities of the bulge and disk lenses, respectively, while the black line represents the full sample. The metallicities of our bulge lenses were estimated using the MDF measured by \citet{2013A&A...549A.147B} from microlensed bulge dwarfs and we estimate the metallicities of our disk lenses using the Galactic metallicity gradients measured by \citet{2013arXiv1311.4569H} from the SDSS-III APOGEE experiment. The bottom panel shows the cumulative fraction of metallicities for the bulge and disk lenses, as well as for the full sample. We find a median metallicity for our simulated sample of 0.17~dex with a 68\% confidence interval of $-0.23<{\rm [M/H]/dex}<0.41$.
  \label{fig:mlens_metals}}
\end{figure}

The resultant metallicity distribution for our simulated microlensing sample is shown in figure~\ref{fig:mlens_metals}. The blue and red lines represent the metallicities of the bulge and disk lenses, respectively, while the black line shows the distribution of the full sample. The median metallicity of the full sample is 0.17~dex with a 68\% confidence interval of $-0.23<{\rm [M/H]/dex}<0.41$ and a 95\% confidence interval of $-1.0<{\rm [M/H]/dex}<0.54$. While not strictly true, we assume that ${\rm [M/H]}$ traces ${\rm [Fe/H]}$ and adopt these values as the ${\rm [Fe/H]}$ values for our simulated microlensing sample. For comparison, the median metallicity of both the HARPS and CPS M dwarf samples is about ${\rm [Fe/H]}_{\rm med}=-0.1~$dex \citep[][MB14]{2013A&A...551A..36N}. As expected, we find that the distribution of metallicities for our simulated microlensing sample is systematically higher than that of RV surveys.

\begin{figure}[t!]
\epsscale{1.1}
\plotone{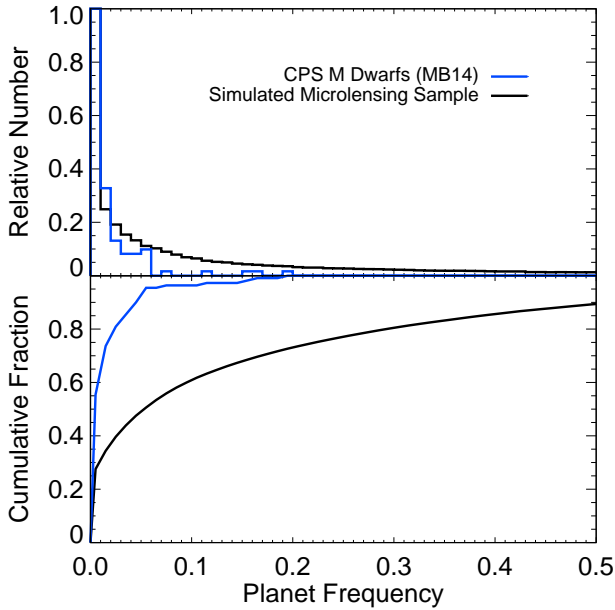}
\caption{The top panel shows the relative number of stars with a frequency of planets with masses $1<m_p\sin{i}/M_{\rm Jup}<13$ on orbits within $a<20~$AU implied by equation~(\ref{eqn:mb_metallicity_corr}) given their mass and metallicity. The blue and black lines represent the planet frequencies for the CPS M dwarf sample and our simulated microlensing sample (which has the same stellar, i.e. lens, mass distribution), respectively. The bottom panel shows the cumulative fraction of systems with a given planet frequency for both samples. We find a mean giant planet occurrence rate of 0.36 for our simulated microlensing sample implied by the MB14 relation, which is a median factor of 13 ($4.4-44$ at 95\% confidence) larger than the value we derive from microlensing and RV constraints from HARPS in \S~\ref{sec:synthesis}.
  \label{fig:rv_mlens_freqs}}
\end{figure}

Now that we have metallicities for the microlensing sample, we compute the frequency of Jupiters and super-Jupiters on orbits within $a<20~$AU implied from the CPS results (MB14) given by equation~(\ref{eqn:mb_metallicity_corr}) using the median values of the fit parameters (i.e. the median normalization and scalings with host mass and metallicity reported by MB14). Figure \ref{fig:rv_mlens_freqs} shows the resultant distributions and cumulative distribution functions of the implied frequency of planets with $1<m_p\sin{i}/M_{\rm Jup}<13$ for our simulated microlensing sample and the CPS sample.

We find a mean occurrence rate of Jupiters and super-Jupiters of $0.36$ is expected for our simulated microlensing sample from the MB14 relation. This expected frequency is discrepant by a median factor of 13 ($4.4-44$ at 95\% confidence) from the actual value. In \S~\ref{sec:synthesis} we derive planet frequencies from the combined constraints of real microlensing surveys and the HARPS RV survey, and from these combined constraints, we find a frequency of planets with masses $1<m_p\sin{i}/M_{\rm Jup}<13$ and periods $1\leq P/{\rm days}\leq 10^5$ of $0.032^{+0.014}_{-0.017}$, consistent with the value reported by MB14 for the CPS M dwarfs of $0.065\pm 0.030$ but still a median factor 2.3 ($0.22-8.8$ at 95\% confidence) times smaller (see previous section for discussion). We will discuss a few possible reasons for the inconsistency in the frequencies implied by the MB14 relation for our simulated microlensing sample and the actual value we find from the combined constraints of microlensing and RV surveys.

First, we pose a question. What if giant planets do not form around bulge stars \citep[e.g.][]{2013MNRAS.431...63T}? To investigate this possibility, we repeat the calculation described above for our simulated microlensing sample, except we set the frequency of giant planets for bulge hosts to zero. We find a mean occurrence rate of 0.25 implied by equation~\ref{eqn:mb_metallicity_corr}, which is discrepant from the actual value by a median factor of 9.1 ($3.0-30$ at 95\% confidence). Thus, while this hypothesis shifts the implied frequency for the microlensing sample in the right direction, it does not seem to be enough to cause agreement. On the other hand, this idea is attractive for another reason. It could also explain the difference in the lens distance, $D_l$, distributions between our simulated sample (which assumes planets are equally common around stars regardless of their location) and that of the actual GA10 microlensing sample. GA10 find a median lens distance of 3.4~kpc, while our simulated microlensing sample yields a median value of 6.7~kpc. If there are no planets in the bulge, then the median distance to planet hosting lenses in the disk is 5.8~kpc. Thus, the idea that planets do not form around bulge stars could help to explain the shorter lens distances inferred from the GA10 sample relative to that inferred from our simulated sample, while leaving the distributions of the other microlensing parameters for these samples in agreement (see \S~5.3.3 of \citet{clanton_gaudi14a} for more information on the properties of our simulated sample and how they compare with those of the GA10 sample).

Next, we examine the possibility that the MB14 relation is not correct. We also compute the implied occurrence rates from the relations between planet frequency and mass and metallicity derived in JJ10 and \citet{2013A&A...551A..36N}. The mean occurrence rates implied by equation~\ref{eqn:jj_metallicity_corr} (JJ10) for our simulated microlensing sample is 0.058, a median factor of just 2.1 ($0.70-7.0$ at 95\% confidence) larger. The mean occurrence rate implied by equation~\ref{eqn:harps_metallicity_corr} \citep{2013A&A...551A..36N} is 0.065, while the value from equation~\ref{eqn:harps_cps_metallicity_corr} \citep{2013A&A...551A..36N} is 0.23. These implied frequencies are median factors of 2.4 ($0.79-7.9$ at 95\% confidence) and 8.3 ($2.8-28$ at 95\% confidence), respectively, larger than the actual value. The more shallow scalings of these other relations do bring the RV-expected planet frequencies closer to agreement. Perhaps surprisingly, the JJ10 relation provides the best agreement between the CPS sample and our simulated microlensing sample, even though their stellar sample included higher mass (FGK) stars whereas the MB14 and \citet{2013A&A...551A..36N} samples include only M dwarfs.

Given the fact that the median metallicity of our simulated microlensing sample is about 0.17~dex (assuming the distribution we derive is correct), while that of the CPS M dwarfs is -0.1~dex, then it seems that the frequency of giant planets must have a weaker dependence on ${\rm [Fe/H]}$ then implied by MB14. Another possibility is that the metallicity dependence saturates at some value, with (e.g.) a flat distribution for metallicities above the saturation value. It could even be a combination of the various effects we discuss, i.e. a suppression of the formation of giant planets in the bulge, a slightly weaker scaling with metallicity, and a saturation of the giant planet frequency above some threshold ${\rm [Fe/H]}$.

Finally, we also note that it is unlikely the MB14 relation between the frequency of planets with masses $1\lesssim m_p\sin{i}/M_{\rm Jup}\lesssim 13$ on orbits with $a<20~$AU extends down to giant planets with masses between $0.1\lesssim m_p\sin{i}/M_{\rm Jup}\lesssim 1$ given the fact that RV surveys generally do not detect the bulk of this planet population due to the steep planetary mass function inferred from microlensing \citet{2010ApJ...710.1641S}. We show in the next section that the frequency of giant planets with masses between $0.1\lesssim m_p\sin{i}/M_{\rm Jup}\lesssim 30$ and periods between $1\leq P/{\rm days}\leq 10^5$ is $f_{\rm G}=0.17^{+0.07}_{-0.08}$, which would seem to suggest that the scaling of giant planet frequency with host metallicity is also a function of planetary mass. This is supported by the results of \citet{2013A&A...551A..36N}, which demonstrate that the scaling of planet frequency with host metallicity for Neptunian hosts (as opposed to Jovian hosts) is not only more shallow, but that it possibly even works in the opposite direction (i.e. that planet frequency of Neptunes is anti-correlated with ${\rm [Fe/H]}$), although the latter is not statistically significant when compared against a constant functional form. These authors therefore determine that a constant functional form of $f=0.03\pm 0.01$ is preferred for Neptunian hosts, which is quite different from the relations they find for Jovian hosts given by equations~(\ref{eqn:harps_metallicity_corr}) and (\ref{eqn:harps_cps_metallicity_corr}).

In the end, the scaling of the frequency of giant planets (in particular of Jupiters and super-Jupiters) with stellar metallicity among M dwarfs remains a puzzle. However, we predict that if future RV surveys can begin detecting the bulk of the giant planet population inferred from microlensing, which typically have $K\sim 1~{\rm m~s^{-1}}$ and $P\sim 9~$yr, these planets will be detected around more metal-poor stars. The frequency of Jupiters and super-Jupiters around metal-rich stars is already found to be very high from RV surveys, which implies that the large population of giant planets with $0.1\lesssim m_p\sin{i}/M_{\rm Jup}\lesssim 1$ inferred from microlensing (and not currently detected by RV surveys) would either be detected around stars with lower metallicities or in multi-planet systems around the metal-rich M dwarfs.

\section{Synthesizing Planet Detection Results from Multiple Detection Methods}
\label{sec:synthesis}
We have demonstrated that microlensing predicts consistent numbers of planet detections for the HARPS and CPS M dwarf surveys in the regions of planet parameter space for which there is some overlap. This enables us to synthesize the detection results, i.e. combine the individual constraints, from microlensing and RV surveys to determine planet frequencies across a very wide region of parameter space. We choose to combine the constraints from the microlensing results \citep[GA10;][]{2010ApJ...710.1641S} with those from the RV survey of HARPS (BX13) since their detection limits have been carefully characterized.

\begin{figure*}[h!]
\epsscale{1.1}
\plotone{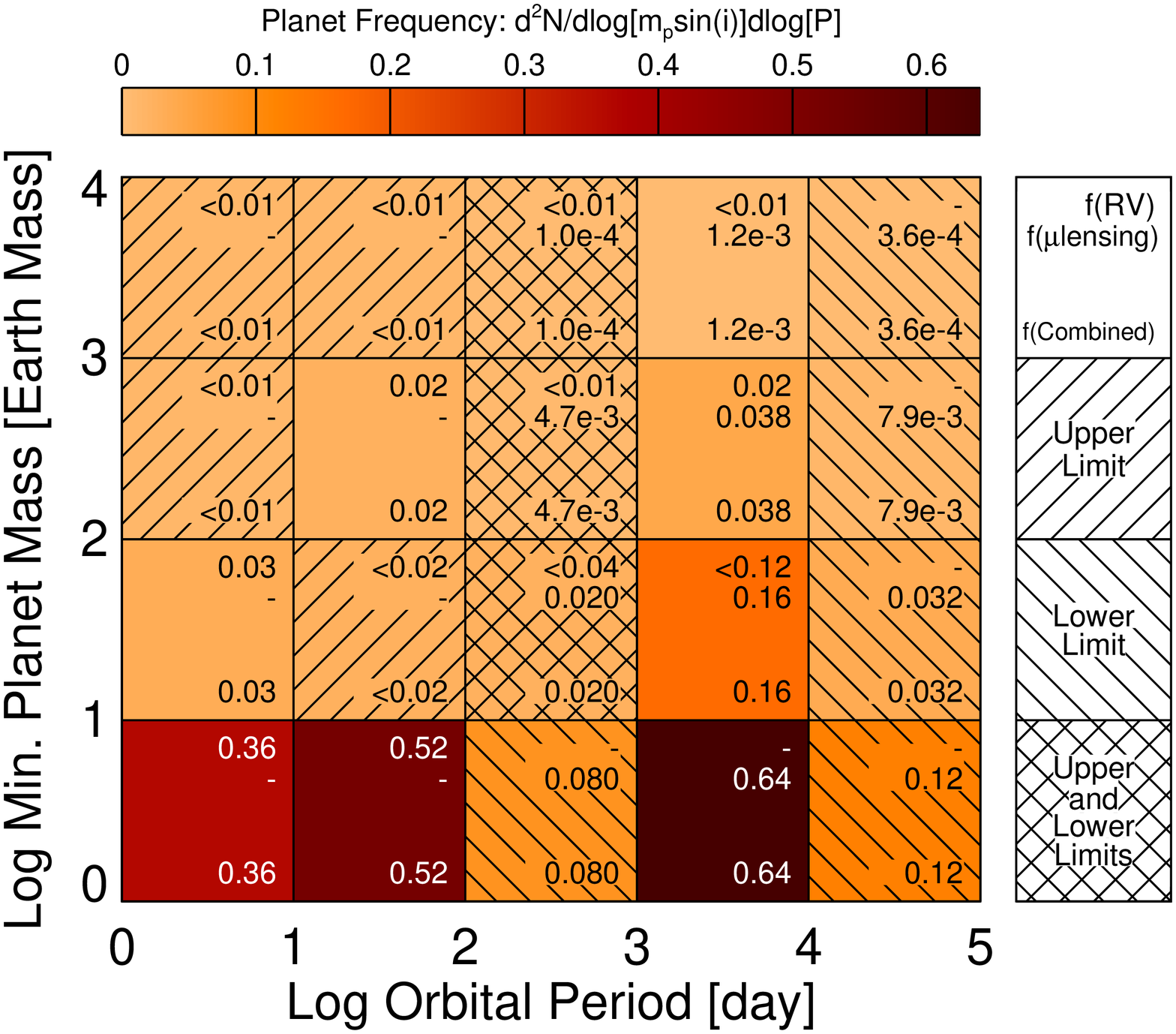}
\caption{Planet frequency as a function of $\log{(P/{\rm day})}$ and $\log{(m_p\sin{i}/M_{\oplus})}$. The numbers displayed in the upper right of each cell are the planet frequencies (or upper limits) derived by BX13 from the HARPS M dwarf sample, while those just under these are the planet frequencies we derive in this study from microlensing. The values in the lower right corner of each cell are the synthesized planet frequencies from both the RV and microlensing constraints (see text for an explanation of how we combine the statistics). The cells are color coded according to the synthesized planet frequency in the corresponding area. In cells where we have a lower limit from microlensing, the color represents this lower limit, whereas the cells that have only upper limits are given colors equal to the quoted 1$\sigma$ upper limits. The uncertainties on all these values are listed in table~\ref{tab:synthesized_fs}.
  \label{fig:freq_plot}}
\end{figure*}

Table \ref{tab:synthesized_fs} and figure \ref{fig:freq_plot} display these combined constraints in bins of $\log{P}$ and $m_p\sin{i}$. Our methods for combining these results are as follows.
\begin{itemize}
\item $1\leq P/{\rm days}\leq 10^2$: Since microlensing has basically no sensitivity to planets with periods $\lesssim 10^2~$days, we use the constraints on the planet frequencies for these periods from the HARPS survey alone.
\item $10^2\leq P/{\rm days}\leq 10^3$: We include constraints from both microlensing and RV within this period range. However, in these bins, BX13 measure only upper limits on the planet frequency. The frequencies we derive in this paper from microlensing are consistent with these upper limits, however they serve as lower limits on the planet frequency in these bins since microlensing surveys are incomplete for these periods. Therefore, the true frequency is likely somewhere between the RV and microlensing estimates in these bins.
\item $10^2\leq P/{\rm days}\leq 10^3$: Although there is some overlap for these periods, this parameter space is dominated by microlensing, and so we adopt the microlensing estimates.
\item $10^4\leq P/{\rm days} \leq 10^5$: The HARPS survey has no sensitivity to these orbital periods (other than trends), while the sensitivity of microlensing surveys cuts off near the short end of this range for higher planet masses. Thus, we adopt the microlensing estimates in these bins but note that, due to the rapidly declining sensitivity of microlensing surveys in this period range (and especially for the lower-mass bins at these periods), the frequencies in these bins are really lower limits.
\end{itemize}
The microlensing results constrain the frequency of planets with projected separations near the Einstein ring for masses down to about $\sim 5~M_{\oplus}$, so the planet frequencies we derive in the mass range $1\leq m_p\sin{i}/M_{\oplus}\leq 10$ from microlensing requires an extrapolation of our mass function (see equation~\ref{eqn:planetary_mass_function}). However, the required extrapolation is only about 0.7~dex in $\log{q}$ for a primary mass of $M_l\sim 0.5~M_{\odot}$.

The planet frequency as a function of $\log{(m_p\sin{i}/M_{\oplus})}$ and $\log{(P/{\rm days})}$ that we derive with the above rules are displayed in figure~\ref{fig:freq_plot}. The cells are color coded according to the synthesized planet frequency in the corresponding area. In cells where we have a lower limit from microlensing, the color represents this lower limit, whereas the cells that have only upper limits are given colors equal to the quoted 1$\sigma$ upper limits.

\begin{table*}
\centering
\caption{\label{tab:synthesized_fs} Planet frequency as measured by RV, $f_{\rm RV}$, and microlensing, $f_{\rm \mu lens}$, surveys. The quantity $f_{\rm syn}$ is the planet frequency derived from constraints on either one or both of RV and microlensing detection results, depending on the sensitivity of these techniques in a given mass and period bin (see text for more details).}
\begin{tabular}{l|rrrrr}
\hline \hline
\multicolumn{1}{c|}{$m_{\rm p}\sin{i}$}  &  \multicolumn{5}{c}{Orbital Period [days]} \\
\multicolumn{1}{c|}{[M$_{\oplus}$]} &  1$-$10  & $10-10^2$  & $10^2-10^3$  & $10^3-10^4$ & $10^4-10^5$\\
        \hline
 & $f_{\rm RV} < 0.01$& $f_{\rm RV} < 0.01$& $f_{\rm RV} < 0.01$& $f_{\rm RV} < 0.01$& $-$\\
$10^3-10^4$& $-$& $-$& $f_{\rm \mu lens} = (1.0^{+1.1}_{-1.0})E-4$& $f_{\rm \mu lens} = (1.2^{+8.4}_{-1.0})E-3$& $f_{\rm \mu lens} = (3.6^{+3.4}_{-3.5})E-4$\\
 & $f_{\rm syn} < 0.01$& $f_{\rm syn} < 0.01$& $(1.0^{+1.1}_{-1.0})E-4 \leq f_{\rm syn} < 0.01$& $f_{\rm syn} = (1.2^{+8.4}_{-1.0})E-3$& $f_{\rm syn} \geq (3.6^{+3.4}_{-3.5})E-4$\\
        \hline
 & $f_{\rm RV} < 0.01$& $f_{\rm RV} = 0.02_{-0.01}^{+0.03}$& $f_{\rm RV} < 0.01$& $f_{\rm RV} = 0.019_{-0.015}^{+0.043}$& $-$\\
$10^2-10^3$& $-$& $-$& $f_{\rm \mu lens} = (4.7^{+3.1}_{-3.4})E-3$& $f_{\rm \mu lens} = 0.038^{+0.023}_{-0.026}$& $f_{\rm \mu lens} = (7.9^{+4.8}_{-5.4})E-3$\\
 & $f_{\rm syn}  < 0.01$& $f_{\rm syn} = 0.02_{-0.01}^{+0.03}$& $(4.7^{+3.1}_{-3.4})E-3\leq f_{\rm syn} < 0.01$& $f_{\rm syn} = 0.038^{+0.023}_{-0.026}$& $f_{\rm syn} \geq (7.9^{+4.8}_{-5.4})E-3$\\
        \hline
 & $f_{\rm RV} = 0.03_{-0.01}^{+0.04}$& $f_{\rm RV} < 0.02$& $f_{\rm RV} < 0.04$& $f_{\rm RV} < 0.12$& $-$\\
$10-10^2$& $-$& $-$& $f_{\rm \mu lens} = 0.020\pm 0.009$& $f_{\rm \mu lens} = 0.16^{+0.068}_{-0.072}$& $f_{\rm \mu lens} = 0.032^{+0.012}_{-0.014}$\\
 & $f_{\rm syn} = 0.03_{-0.01}^{+0.04}$& $f_{\rm syn}  = < 0.02$& $0.020\pm 0.009 \leq f_{\rm syn} < 0.04$& $f_{\rm syn} = 0.16^{+0.068}_{-0.072}$& $f_{\rm syn} \geq 0.032^{+0.012}_{-0.014}$\\
        \hline
 & $f_{\rm RV} = 0.36_{-0.10}^{+0.24}$& $f_{\rm RV} = 0.52_{-0.16}^{+0.50}$& $-$& $-$& $-$\\
$1-10$& $-$& $-$& $f_{\rm \mu lens} = 0.080\pm 0.031$& $f_{\rm \mu lens} = 0.64^{+0.25}_{-0.26}$& $f_{\rm \mu lens} = 0.12^{+0.051}_{-0.049}$\\
 & $f_{\rm syn} = 0.36_{-0.10}^{+0.24}$& $f_{\rm syn} = 0.52_{-0.16}^{+0.50}$& $f_{\rm syn} \geq 0.080\pm 0.031$& $f_{\rm syn} = 0.64^{+0.25}_{-0.26}$& $f_{\rm syn} \geq 0.12^{+0.051}_{-0.049}$\\
\hline\hline
\end{tabular}
\end{table*}

We also derive the integrated frequencies of various populations of planets. In order to compute the contribution due to microlensing constraints, we bin the output of our simulations over the appropriate area of planet mass and period, and add the result with the contribution from the RV constraints of the HARPS survey in accordance with the rules we list above. However, because BX13 report the planet frequencies in decade bins of mass and period, we cannot robustly determine true planet frequencies from their survey for planet populations with mass cutoffs that are not equal to 1, 10, 100, or 1000 $M_{\oplus}$. To determine their frequencies, BX13 compute an effective number of stars whose detection limits confidently exclude the existence of planets with similar masses and periods. This effective number of stars is only determined in specific bins and we do not know how these are distributed within a given bin. Thus, we must make approximations and we compute the frequencies in the following manner. We assume the mean number of planets of a given population (e.g. giant planets) is equal to the actual number BX13 detect, drawing a value from a Poisson distribution with such a mean. We then divide by our own effective number of stars which we calculate by normalizing the frequency of the population to a specific value at the actual number of detections. This specific value is an approximation of the planet frequency which we compute by assuming the planet frequency is evenly distributed throughout the BX13 decade bins. We numerically determine uncertainties for the HARPS constraints simply from the Poisson error on the number of detections, combining them (numerically) with the uncertainties from the microlensing contribution.

We find the frequency of giant planets with $30\lesssim m_p\sin{i}/M_{\oplus}\lesssim 10^4$ and $1\leq P/{\rm days}\leq 10^4$ to be $f_{\rm G}=0.15^{+0.06}_{-0.07}$, or $f_{\rm G}=0.17^{+0.07}_{-0.08}$ over the period interval $1\leq P/{\rm days}\leq 10^5$. A more conservative definition of giant planets ($50\lesssim m_p\sin{i}/M_{\oplus}\lesssim 10^4$), yields a frequency of $f_{\rm G'}=0.11\pm 0.05$ ($1\leq P/{\rm days}\leq 10^4$) or $f_{\rm G'}=0.13\pm 0.06$ ($1\leq P/{\rm days}\leq 10^5$). The frequency of Jupiters and super-Jupiters ($1\lesssim m_p\sin{i}/M_{\rm Jup}\lesssim 13$) with periods $1\leq P/{\rm days}\leq 10^4$ is $f_{\rm J}=0.029^{+0.013}_{-0.014}$ or $f_{\rm J}=0.032^{+0.014}_{-0.017}$ over the period interval $1\leq P/{\rm days}\leq 10^5$, consistent within $1\sigma$ of the measurement by MB14 from the CPS M dwarfs of $f_{\rm J}=0.065\pm 0.030$. As we mentioned in \S~{\ref{subsec:johnson_comparison}}, although this frequency is consistent with that of MB14, it is nevertheless a median factor of 2.3 ($0.22-8.8$ at 95\% confidence) times smaller, potentially due to the fact that microlensing is missing a population of very long-period super-Jupiters that is being inferred by MB14. Integrating over the entire mass range, we find the frequency of all planets with $1\leq m_p\sin{i}/M_{\oplus}\leq 10^4$ and $1\leq P/{\rm days}\leq 10^4$ to be $f_p=1.9\pm 0.5$ or $f_p=2.0\pm 0.5$ over the period interval $1\leq P/{\rm days}\leq 10^5$.

Microlensing surveys are sensitive to planets at projected separations out to roughly $s=2.5$. For the typical lens star ($M_l\sim 0.5~M_{\odot}$), this corresponds to $r_{\perp}\sim sR_E\sim 7$~AU, or $P\sim 10^4~$days (assuming $D_l/D_s=1/2$ and $a=r_{\perp}$). Thus, while there is some sensitivity to planets beyond $10^4~$days, we are not able to derive strong constraints on planet frequencies for periods beyond $10^4$ days. In this paper, we restrict our integrated estimates of planet frequency to periods $P\leq 10^4~$days. In a future paper, we plan to more accurately characterize the giant planet frequency at longer periods by including constraints from direct imaging surveys.

We note that the median number of epochs for stars in the HARPS M dwarf sample (BX13) is 8, with only 14 of their total 97 stars having more than 40 total epochs. Of these 14 stars, seven are planet hosts (Gl 176, Gl 433, Gl 581, Gl 667C, Gl 674, Gl 832, Gl 876), three show periodic variability that BX13 show to correlate with stellar activity (Gl 205, Gl 388, Gl 479), three are shown to have statistically significant long-term RV trends (Gl 1, Gl 273, Gl 887), and one is a bright M2 star of which BX13 take exposures to construct a numerical weighted mask to cross-correlate their spectra and compute RVs (Gl 877). There is only one planet host in their sample with less than 40 total epochs (Gl 849 with $N=35$), but which was previously known to host a giant planet \citet{2006PASP..118.1685B}. If it is the case that the planet hosts were specifically targeted for additional observations as a result of the presence of a planet, and these additional observations were not excluded when quantifying the planet sensitivity, then the planet frequencies inferred from these data are biased.  However, we are unable to quantify the magnitude, or even the sign, of this bias.

\subsection{Comparison of Combined Constraints with Other Measurements of Planet Frequency}
\label{subsec:freq_comparisons}
Now that we have derived the planet frequency around M dwarfs across a very wide region of planet parameter space, we can compare with other measurements of planet frequency. In particular, we make rough comparisons with frequencies from a sample of M dwarfs from \emph{Kepler} by \citet{2013ApJ...767...95D} and by \citet{2013ApJ...764..105S}, as well as a measurement of the giant planet ($0.3\leq m_p\sin{i}/M_{\rm Jup}\leq 15$) frequency around F, G, and K dwarfs by \citet{2008PASP..120..531C}.

\subsubsection{\emph{Kepler} M Dwarfs}
\label{subsubsec:kep_comparison}
\citet{2013ApJ...767...95D} refine the stellar parameters of a sample of M dwarfs from \emph{Kepler} and compute planet frequencies as a function of orbital period and planetary radius. We perform a rough comparison with their results in the bins corresponding to planet masses between $1\leq m_p\sin{i}/M_{\oplus}\leq 10$ and orbital periods between $1\leq P/{\rm days}\leq 10^2$. The empirical mass-radius relations derived in \citet{2014ApJ...783L...6W} tell us that a planetary radius of $\approx 4~R_{\oplus}$ corresponds to a mass of $\approx 10~M_{\oplus}$. Assuming planetary densities identical to that of the Earth, the \citet{2014ApJ...783L...6W} relations say $1~R_{\oplus}$ corresponds to $1~M_{\oplus}$. Thus, in terms of radius, we compare the frequencies derived by \citet{2013ApJ...767...95D} between $1\leq R_p/R_{\oplus}\leq 4$ for periods between $1\leq P/{\rm days}\leq 10^2$ (ignoring the $\sin{i}$ factor). We choose to compare results in these specific bins of mass (radius) and orbital period because these are the regions of this parameter space for which we expect the most overlap between \emph{Kepler} and the HARPS RV survey, from which we derive our constraints on planet frequency in these bins.

We rebin the data in figure~15 of \citet{2013ApJ...767...95D}, adding up their planet frequencies in the bins between $1\leq R_p/R_{\oplus}\leq 4$ and $1\leq P/{\rm days}\leq 10$ (and multiplying by the appropriate fractions for the cells that are not fully contained in this range, assuming a uniform distribution in $\log{R}$ and $\log{P}$). This roughly yields the frequency of planets with masses between $1\leq m_p\sin{i}/M_{\oplus}\leq 10$ in the same period range. We find this frequency to be $0.23\pm 0.03$, which is nearly consistent with the frequency inferred from the HARPS survey by BX13 of $0.36^{+0.50}_{-0.10}$. In this same mass interval but for orbital periods of $10\leq P/{\rm days}\leq 10^2$, we compute a frequency from the \citet{2013ApJ...767...95D} results of $0.51\pm 0.10$. The frequency in the corresponding bin measured by BX13 is $0.52^{+0.50}_{-0.16}$, consistent with our calculation from the \emph{Kepler} M dwarfs.

According to the \citet{2014ApJ...783L...6W} mass-radius relation, a $4~R_{\oplus}$ planet will have a mass of about $25~M_{\oplus}$. As we discussed in \S~\ref{sec:giant_planet_def}, the lowest mass of a ``giant planet'' is uncertain, and likely encompasses a range of masses. Even more uncertain, then, is the transition radius between rocky, icy, and giant planets. However, if we make the simple assumption that all the planets with radii $R_p>4~R_{\oplus}$ in the \citet{2013ApJ...767...95D} M dwarf sample from \emph{Kepler} are giant planets, then we calculate the frequency of giant planets with masses $m_p\sin{i}\gtrsim 30~M_{\oplus}$ and periods $1\leq P/{\rm days}\leq 10$ to be $0.014\pm 0.007$. This is nearly consistent with the frequency of planets in the same mass and period ranges measured BX13 of $0.043\pm 0.021$.

\citet{2013ApJ...764..105S}, assuming the five planets of Kepler-32 \citep{2012ApJ...750..114F} are representative of the full ensemble of planet candidates orbiting the \emph{Kepler} M dwarfs, infer a planet occurrence rate of $1.0\pm 0.1$ planet per star. While \citet{2013ApJ...764..105S} do not explicitly state the planetary radius and orbital period intervals over which this measurement is integrated, examining their figure~6 seems to indicate intervals of $m_p\gtrsim1~M_{\oplus}$ and $P\lesssim150~{\rm days}$, where they have adopted the planetary mass-radius relation of \citet{2011ApJS..197....8L} which takes the form $m_p\propto R_p^{2.06}$. In these same intervals, we find an occurrence rate of $0.94^{+0.35}_{-0.26}$, consistent with \citet{2013ApJ...764..105S}. \citet{2013ApJ...764..105S} also calculate the occurrence rate of planets with $R_p>2~R_{\oplus}$ (corresponding to a mass of $\approx 5~M_{\oplus}$ according to the mass-radius relation of \citealt{2014ApJ...783L...6W}) and $P<50~{\rm days}$ to be $0.26\pm 0.05$. In these same intervals, we again find a consistent planet frequency of $0.37^{+0.18}_{-0.13}$.

\subsubsection{Planet Frequency Around FGK Dwarfs}
\label{subsubsec:kep_comparison}
\citet{2008PASP..120..531C} analyzed a sample of RV-monitored FGK stars and measured the occurrence rate of planets with masses between $0.3\leq m_p\sin{i}/M_{\rm Jup}\leq 10$ and orbital periods $P<5.2~$yr to be $0.085\pm 0.013$. They extrapolate to find the frequency of such planets with orbital semimajor axes $a<20~$AU, assuming either a flat distribution in $P$ beyond $2000~$days or a power-law distribution ($\propto P^{0.26}$), to be $0.17\pm0.03$ and $0.19\pm 0.03$, respectively. Around a Solar-type star, 20~AU is roughly 7.4 times the location of the ice line, assuming $a_{\rm ice}=2.7~{\rm AU}(M/M_{\odot})^{2/3}$, where we have adopted the scaling found by \citet{2008ApJ...673..502K} for mass accretion rates that are proportional to stellar mass, $\dot{M}\propto M_{\star}$. In order to compare planet frequencies between FGK and M dwarf populations, we want to examine orbital separations that probe similar formation environments, so we compute the frequency from our combined constraints over the same range of planetary masses, but for orbital separations that are within 7.4 times the location of the ice line, which is about 12.5~AU for the typical microlensing star of $0.5~M_{\odot}$. Thus, we find for masses in the range $0.3\leq m_p\sin{i}/M_{\rm Jup}\leq 10$ and $a<12.5~$AU ($P<62.5~$yr) a frequency of $0.072^{+0.034}_{-0.038}$, which is a median factor of 2.8 ($0.81-9.5$ at 95\% confidence) to 3.1 ($0.89-9.9$ at 95\% confidence) times smaller than, and thus marginally inconsistent at the $2\sigma$ level with, the values found by \citet{2008PASP..120..531C} for FGK stars.

If we extrapolate the \citet{2008PASP..120..531C} planetary mass function to include all giant planets ($0.1\leq m_p\sin{i}/M_{\rm Jup}\leq 10$) within $a<20~$AU, we find a frequency of $0.31\pm 0.07$. Our combined constraints give a giant planet frequency for $0.1\leq m_p\sin{i}/M_{\rm Jup}\leq 10$ and $P<62.5~$yr of $0.16\pm 0.07$. This is a median factor of 2.2 ($0.73-5.9$ at 95\% confidence) times smaller than that which we calculate by extrapolating the \citep{2008PASP..120..531C} result. Thus, while giant planets are not intrinsically rare around M dwarfs, they are rarer than the population observed around FGK dwarfs at $1\sigma$ and marginally inconsistent at the $2\sigma$ level.

\section{Summary and Discussion}
\label{sec:discussion}
In this paper, we map the observable parameters ($q,s$) of the population of planets inferred from microlensing into the observables ($K,P$) of an analogous population of planets orbiting a stellar sample monitored with RV. We derive joint distributions of these RV observables for simulated samples of microlensing systems with similar stellar mass distributions as the M dwarf RV surveys of HARPS (BX13) and CPS (MB14). We then apply the actual RV detection limits reported by BX13 to predict the number of planet detections and long-term RV trends we expect the HARPS survey to find, and we apply roughly estimated detection limits to make predictions for the CPS sample. Comparing our predictions with the actual numbers reported by these RV surveys, we find consistency. We predict that HARPS should find $N_{\rm det}=1.4\pm 0.8$ planets right at the edge of their survey limit, and indeed, they find one such planet around Gl 849 (BX13). This star also appears in the CPS sample, where this very same planet was originally discovered \citep{2006PASP..118.1685B}. We expect the CPS survey to detect $N_{\rm det}=4.7^{+2.5}_{-2.8}$ planets with periods $P\gtrsim 100~$days and masses $m_p\sin{i}\gtrsim 10^2~M_{\oplus}$. The number of such planets they actually detect is four, around the stars Gl 179, Gl 317, Gl 649, and Gl 849 \citep{2010ApJ...721.1467H,2007ApJ...670..833J,2010PASP..122..149J,2006PASP..118.1685B}.

The fact that our predicted numbers of detections and the actual numbers are consistent implies that microlensing and RV surveys are largely disjoint, with only a small amount of overlap for orbital periods between roughly $100-10^3~$days and planetary masses larger than about a Jupiter mass. This limited overlap is such that, due to the steeply declining planetary mass function, RV surveys infer low giant planet frequencies around M dwarfs, detecting only the high-mass end of the giant planet population ($m_p\gtrsim M_{\rm Jup}$) inferred by microlensing. For RV surveys to be sensitive to the majority of this population, measurement precisions of $\sim 1~{\rm m~s^{-1}}$ (including instrumental errors and stellar jitter) over time baselines of $\sim 10~$years are required. The frequency of Jupiters and super-Jupiters around metal-rich stars is already found to be very high from current RV surveys, which implies that the large population of giant planets with $0.1\lesssim m_p\sin{i}/M_{\rm Jup}\lesssim 1$ inferred from microlensing (and not currently detected by RV surveys) would either be detected by future, more sensitive RV surveys around stars with lower metallicities or in multi-planet systems around the metal-rich M dwarfs.

However, we are left with a puzzle concerning the scaling of the frequency of Jovian planets ($1\lesssim m_p\sin{i}/M_{\rm Jup}\lesssim 13$) with stellar metallicity inferred from the CPS M dwarf sample \citep{2014ApJ...781...28M}. We estimate the metallicity distribution of our simulated microlensing sample using the bulge MDF of \citet{2013A&A...549A.147B} and the Galactic metallicity gradients from \citet{2013arXiv1311.4569H} and find a median metallicity of ${\rm [M/H]} = 0.17~$dex with a 68\% confidence interval of $-0.23<{\rm [M/H]}/{\rm dec}<0.41$. Using this metallicity distribution, we find that the occurrence rate implied by the scaling inferred by MB14 is over-predicted by a median factor of 13 ($4.4-44$ at 95\% confidence) relative to the actual frequency found by microlensing surveys. This could suggest that the MB14 relation is incorrect or perhaps incomplete. A significantly shallower scaling with metallicity seems to be required for agreement (more in line with that reported by \citealt{2010PASP..122..905J} or perhaps \citealt{2013A&A...551A..36N}), or perhaps the metallicity dependence saturates at some value, with (e.g.) a flat distribution for metallicities above the saturation value. We also investigate another possibility. What if giant planets do not form around bulge stars \citep[e.g.][]{2013MNRAS.431...63T}? We show that if this were true, the occurrence rate for the microlensing sample implied by the MB14 relation moves closer to agreement with the measured value (a median factor of 9.1, or $3.0-30$ at 95\% confidence, discrepant), but probably does not account for the full difference. This solution would also be attractive because it could partially explain the difference in the lens distance distributions between our simulated microlensing sample and the GA10 sample.

We also point out that it seems unlikely the relations between planet frequency and metallicity hold for giant planets with masses $0.1\lesssim m_p\sin{i}/M_{\rm Jup}\lesssim 1$ given the fact that RV surveys are not sensitive to the bulk of the giant planet population inferred from microlensing surveys. This suggests that the scaling of giant planet frequency with host metallicity is a function of planetary mass. This hypothesis is supported by the results of \citet{2013A&A...551A..36N}, which suggest that the scaling of planet frequency with host metallicity is significantly different between Jovian and Neptunian hosts.

Finally, since we have demonstrated that the giant planet frequencies measured by microlensing and RV surveys are actually consistent, we are able to combine their constraints to determine planet frequencies across a very wide region of parameter space. The combined constraints on the giant planet occurrence rate around M dwarfs as a function of orbital period and planet mass are summarized in table~\ref{tab:synthesized_fs} and plotted in figure~\ref{fig:freq_plot}. We also show that the planet frequencies in the mass range $1\leq m_p\sin{i}/M_{\oplus}\leq 10$ and period range $1\leq P/{\rm days}\leq 10^2$ are consistent with the detection results from the \emph{Kepler} M dwarf sample reported by \citet{2013ApJ...767...95D} and \citet{2013ApJ...764..105S}. We can integrate over various regions of this plane to compute total planet frequencies. 

We find the frequency of giant planets with $30\lesssim m_p\sin{i}/M_{\oplus}\lesssim 10^4$ and $1\leq P/{\rm days}\leq 10^4$ to be $f_{\rm G}=0.15^{+0.06}_{-0.07}$. For a more conservative definition of giant planets ($50\lesssim m_p\sin{i}/M_{\oplus}\lesssim 10^4$), we find $f_{\rm G'}=0.11\pm 0.05$. The frequency of Jupiters and super-Jupiters ($1\lesssim m_p\sin{i}/M_{\rm Jup}\lesssim 13$) with periods $1\leq P/{\rm days}\leq 10^4$ is $f_{\rm J}=0.029^{+0.013}_{-0.015}$, consistent with the measurement by MB14 of $f_{\rm J}=0.065\pm 0.030$. We find the frequency of all planets with $1\leq m_p\sin{i}/M_{\oplus}\leq 10^4$ and $1\leq P/{\rm days}\leq10^4$ to be $f_p=1.9\pm 0.5$. These planet frequencies are closer to lower limits on the planet frequency, because our combined constraints on the planet frequency include the lower limits in the period range $10^2-10^3~$days, where the sensitivity of microlensing surveys declines.

This is a very broad result, covering four orders of magnitude in planetary mass and four orders of magnitude in orbital period. But perhaps more importantly, it demonstrates that it is possible to get a more complete picture of the demographics of exoplanets by including constraints from multiple discovery methods. In a future paper, we plan to compare and synthesize the planet detection results found here with those from direct imaging surveys.

\acknowledgments
This research has made use of NASA's Astrophysics Data System and was partially supported by NSF CAREER Grant AST-1056524. We thank John Johnson and Benjamin Montet for helpful comments and conversations.

\appendix
\section{Properties of Predicted RV Detections}
\label{sec:appendix}
The expected numbers of planet detections and trends we find for the HARPS and CPS RV surveys are consistent with their reported values. Here, we aim to determine if there is a subset of the planet population inferred from microlensing that we predict RV surveys will preferentially discover. We have already demonstrated that RV surveys are only able to detect the high-mass end of the giant planet population inferred from microlensing, but it would be interesting if, for example, RV surveys were more sensitive to analogs of the microlensing planets found around disk lenses relative to analogs of those found around bulge lenses. Both RV and microlensing techniques have their own selection effects which are imprinted on the distributions of the physical and orbital properties of the planets detectable by both methods and could, in principle, constrain the subset of microlensing planets accessible by RVs (see Section 2 of \citet{clanton_gaudi14a} for a detailed description of key differences between these two techniques). Understanding the conflation of the selection effects from both of these exoplanet discovery methods will improve our understanding of the overlap between microlensing and RV surveys.

\begin{figure}[h!]
\epsscale{0.8}
\plotone{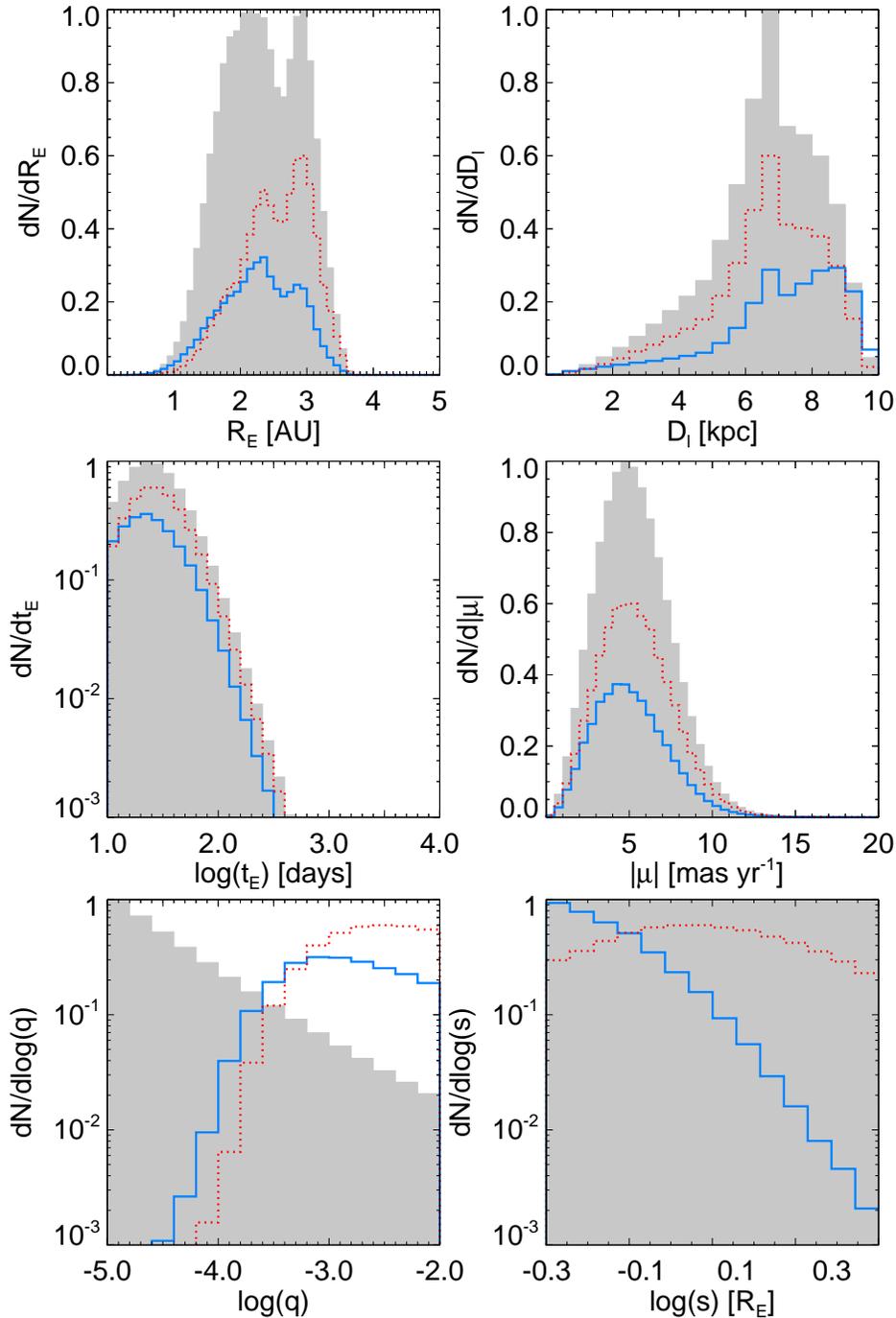}
\caption{Distributions of select microlensing and planetary parameters, distinguishing between detections and trends. In each of the plots, the solid grey distribution represents both the bulge and disk events for all the events falling within $\pm \sigma_{\rm M_{\star}}$ of the mass of each star in the HARPS sample. The blue solid line represents the predicted detections and the red dotted line represents the predicted trends produced by planets orbiting host stars in both the disk and the bulge.
  \label{fig:final_bonfils_mlens_dists_det_trends}}
\end{figure}

Our simulations provide some diagnostic power to examine the properties of the predicted microlensing planets that RV surveys identify as planets and trends. Since we have a much better grasp on the detection sensitivities of the HARPS sample than we do for the CPS sample, we use the results of our comparison with HARPS sample in this section. Figure \ref{fig:final_bonfils_mlens_dists_det_trends} shows distributions of $R_E$, $D_l$, $t_E$, $\left|\boldsymbol{\mu}\right|$, $q$ and $s$ for our simulated planetary microlensing events which we predict BX13 to identify as either a detection or a trend. In order to explain these plots, we need to know how $K$ and $P$ scale with the Einstein radius, $R_E$, the distance to the lens, $D_l$, the distance to the source, $D_s$, the planet/star projected separation, $s$, and the planet/star mass ratio, $q$. The radial velocity of the host star scales as
\begin{equation}
    K \propto \frac{a}{P}q\; ,
\end{equation}
where $q$ is the mass ratio. Combining this with Kepler's 3rd law, we obtain
\begin{equation}
    K \propto a^{-1/2}M_l^{1/2}q\; . \label{eqn:k_propto_a_m_q}
\end{equation}
Given an inclination and mean anomaly, the semi-major axis scales as $a\propto sR_E$. Using this semi-major axis scaling and $R_E\propto M_l^{1/2}\left[x\left(1-x\right)\right]^{1/2}$, where $x\equiv D_l/D_s$, we can substitute the relevant quantities into equation (\ref{eqn:k_propto_a_m_q}) to find the scaling of the velocity of the host/lens star with $R_E$, $s$, $x$, and $q$:
\begin{equation}
    K \propto R_E^{1/2}s^{-1/2}\left[x\left(1-x\right)\right]^{-1/2}q\; . \label{eqn:k_scaling_relation}
\end{equation}
By a similar process, we can start with Kepler's third law and find the scaling of the period with $R_E$, $s$, and $x$ to be
\begin{equation}
    P \propto R_E^{1/2}\left[x\left(1-x\right)\right]^{1/2}s^{3/2}\; . \label{eqn:p_scaling_relation}
\end{equation}

The top left panel of Figure \ref{fig:final_bonfils_mlens_dists_det_trends} shows that the predicted trends tend to have larger Einstein radii than the predicted detections. At fixed $s$, $q$, $D_l$, and $D_s$, equation (\ref{eqn:p_scaling_relation}) becomes $P \propto R_E^{1/2}$, meaning that longer periods correspond to larger Einstein radii. Since one of our criteria for a planet detection includes have a period that is less than the duration of observations, so that a Keplerian fit could be performed with full phase coverage, the planets identified as trends will have periods longer than those identified as detections. Also, at fixed $s$, $q$, $D_l$, and $D_s$, equation (\ref{eqn:k_scaling_relation}) reduces to $K\propto R_E^{1/2}\propto M_l^{1/4}$. In general, RV signals increase with planet mass and decrease with period. Planets identified as trends have larger periods than those identified as detections, so in order to produce a detectable signal (i.e. $K$), trends must necessarily be associated with larger planet masses which requires larger $M_l$ at fixed $q$.

This is consistent with what is seen in the top right and middle left panels of Figure \ref{fig:final_bonfils_mlens_dists_det_trends}. We see that the trends have a preference for longer timescales. At fixed relative proper motions, $\boldsymbol{\mu}$, and lens distances, $D_l$, larger Einstein radii mean longer timescale events because $t_E \propto R_E$. Since disk lenses tend to produce longer timescale events than bulge lenses, Figure \ref{fig:final_bonfils_mlens_dists_det_trends} suggests that trends should be found preferably around disk lenses. Both the detections and trends show non-trivial dependence on the lens distance. For $x \ll 1$, we have from equations (\ref{eqn:k_scaling_relation}) and (\ref{eqn:p_scaling_relation}), the scalings $K \propto D_l^{-1/4}$ and $P \propto D_l^{3/4}$. This means that for lens distances much smaller than the source distance, the RV signal increases with increasing lens distance. For $x \sim 1$, the these scaling relations become $K \propto D_{\rm ls}^{-1/4}$ and $P \propto D_{\rm ls}^{3/4}$, where $D_{\rm ls} \equiv D_s-D_l$, meaning that for lenses close to the source, the RV signal also increases with increasing lens distance. At intermediate lens distances (near $\sim 5~$kpc), there is a relative decrease in RV signals.

The bottom two panels of Figure \ref{fig:final_bonfils_mlens_dists_det_trends} illustrate that predicted trends tend to have larger mass ratios relative to the predicted detections, which follows the fact that larger planet masses produce enhanced RV signals relative to smaller planet masses at fixed $a$ and $M_l$. Predicted trends also tend have larger values of projected separation, producing longer periods, hence why these events are predicted to preferentially show up as trends rather than detections. At fixed $R_E$ and $x$, equation (\ref{eqn:p_scaling_relation}) shows the strong dependence on projected separation, $P\propto s^{3/2}$, for the longer period planets that are identified as trends.

We also investigated how the distributions of microlensing and orbital parameters of the predicted detections and trends depend on their host star locations. This is an important question to answer because it might provide insight into the differences in detections between RV and microlensing surveys and distinguish potential differences in the actual sample of host stars probed by these two methods. We have shown that RV surveys have decreased sensitivity to intermediate host star distances relative to close and far, i.e. close to the source star in the associated microlensing event, host star distances. So, for example, if RV surveys are mostly sensitive to analogs of the microlensing planets in the bulge, it could mean we predict that they would find more planets than what they actually find, if for some reason, planets do not actually form in the bulge \citep[see e.g.][]{2013MNRAS.431...63T}. Additionally, any distance-dependent gradient of properties that affect planet frequency, e.g. the Galactic metallicity gradient \citep[see][]{2012ApJ...746..149C}, could affect our predictions due to the relative RV sensitivity dependence on host star distance. We find that there is not a significant preference for planets around disk or bulge hosts.

\begin{figure}[h!]
\epsscale{0.8}
\plotone{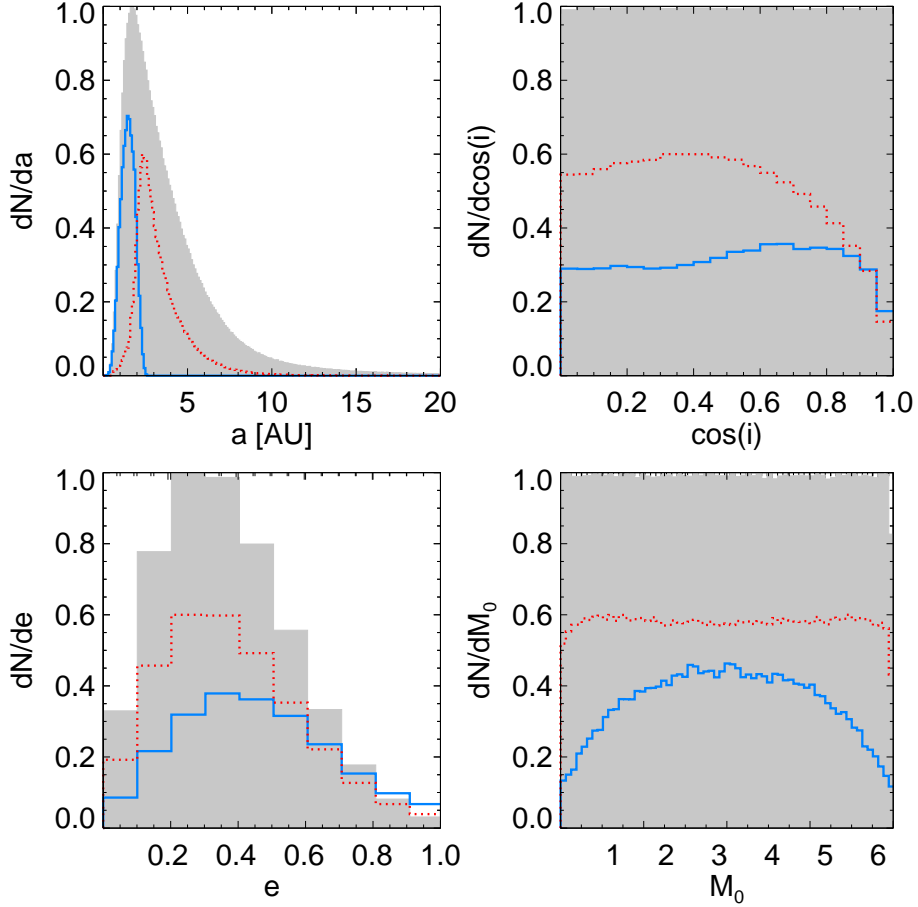}
\caption{Distributions of select orbital parameters associated with the predicted planetary detections and trends, distinguishing between detections and trends. In each of the plots, the solid grey distribution represents both the bulge and disk contributions for all the events falling within $\pm \sigma_{\rm M_{\star}}$ of the mass of each star in the HARPS sample. The blue solid line represents the predicted detections and the red dotted line represents the predicted trends produced by planets orbiting host stars in both the disk and the bulge. There is a pile-up of planets with eccentricities of zero, accounting for 38\% of the population, but these are not drawn in the plot so that the form of the rest of the distribution is visible. The distribution of eccentricities matches that of our prior, except for a slight over-representation of high eccentricity ($e>0.8$) planets counted as detections and trends. In practice, planets with eccentricities $e\gtrsim 0.6$ will not be detectable by RVs \citep{2008PASP..120..531C}, however, these high eccentricity planets only account for a small percentage of the total population.
  \label{fig:final_bonfils_orb_dists_det_trend}}
\end{figure}

Next, we plot the distributions of select orbital parameters ($a$, $\cos{i}$, $e$, $M_0$) of the planets for which we predict RV surveys to identify as either a detection or a trend in figure \ref{fig:final_bonfils_orb_dists_det_trend}. The top left panel shows that the semimajor axis distribution of the predicted detections peaks at a smaller value than that for the predicted trends, and lacks the tail to large separations seen for the trends. The sharp cutoff in the distribution of semimajor axes for the detections is set by the maximum time baseline ($T_{\rm max}\approx 5.75~$yr) in the HARPS sample since we adopt the criterion that $P\leq T$ for a planet detection. In the top right panel, the distributions of $\cos{i}$ for the predicted planet detections and trends show a strong decline as $\cos{i}\rightarrow 1$, corresponding to face-on orbits where RV has no sensitivity since $K\propto \sin{i}$.

The bottom left panel of figure \ref{fig:final_bonfils_orb_dists_det_trend} displays the eccentricity distributions. The planets on circular orbits, which make up 38\% of the population, are not included in this plot to show detail in the rest of the distributions. We find that the eccentricity distribution of both the detections and trends are similar in form to the prior we adopt (see \citet{clanton_gaudi14a}), except for a slight over-representation at high eccentricities, $e>0.8$. These planets would, in practice, be much less detectable with the RV technique relative to those on circular orbits \citep{2008PASP..120..531C}, however these high eccentricity planets only account for a small percentage of the total population. The bottom right panel shows that the distribution of the mean anomalies for the predicted planet detections is slightly peaked at apastron, $M_0 = \pi$. This is due to the fact that planets on eccentric orbits spend more time at apastron where microlensing is more likely to catch them during an event. Additionally, planets detected by microlensing at apastron are near the minimum allowed period given their projected separation, and so tend to have shorter periods (see equation 21 of \citet{clanton_gaudi14a}). This does not seem to be the case for the predicted long-term RV trends.

\clearpage

\bibliographystyle{hapj}
\bibliography{myrefs}
\end{document}